\documentclass{article}
\usepackage[utf8]{inputenc}
\pdfoutput=1
\usepackage{setspace}
\usepackage{palatino}
\usepackage{graphicx}
\usepackage{longtable}
\usepackage{listings}
\usepackage{float}
\usepackage{titling} 
\usepackage{multirow}
\usepackage{lscape}
\usepackage{amsmath}
\usepackage{amssymb}
\usepackage[a4paper, total={6in, 9.5in}]{geometry}
\usepackage{array}
\usepackage[normalem]{ulem}
\fontfamily{ppl}\selectfont 
\usepackage{eqparbox}
\usepackage{arydshln}
\usepackage{mathrsfs}
\usepackage[american]{babel}

\usepackage[figuresleft]{rotating}

\usepackage{bm}
\usepackage{caption}
\usepackage{subcaption}

\usepackage[textsize=small]{todonotes}

\usepackage{bm}
\usepackage{caption}
\usepackage{subcaption}

\newcommand{\appendixqqsection}[1]{\addtocounter{section}{1}
   \setcounter{table}{0}
   \setcounter{figure}{0}
   \setcounter{equation}{0}
   \setcounter{subsection}{0}
  \section*{Supplementary \Alph{section}: #1}
}

\newcommand\appendixqq{
   \setcounter{section}{0}
   \renewcommand\thesection{\Alph{section}}
   \renewcommand{\thesubsection}{\thesection. \arabic{subsection}}
   \renewcommand{\thesubsubsection}{\thesubsection .\arabic{subsubsection}}
   \renewcommand{\thefigure}{\thesection.\arabic{figure}}
   \renewcommand{\thetable}{\thesection.\arabic{table}}
}

\usepackage[natbibapa]{apacite}
\renewcommand{\APACrefDOI}[2]{} 

\PassOptionsToPackage{hyperindex,breaklinks}{hyperref}
\usepackage{hyperref}
\definecolor{darkblue}{rgb}{0, 0, 0.5}
\hypersetup{colorlinks=true,citecolor=darkblue, linkcolor=darkblue, urlcolor=darkblue}

\renewenvironment{abstract}
 {\par\noindent\textbf{\abstractname: }\ \ignorespaces}
 {\par\medskip}

\title{Correspondence analysis: handling cell-wise outliers %by reducing their contribution 
via the reconstitution algorithm}

\author{Qianqian Qi\\
   {\small\raggedright Utrecht University}\\
   \href{mailto:q.qi@uu.nl}{\texttt{q.qi@uu.nl}} 
\and David J. Hessen\\
   {\small\raggedright Utrecht University}\\
   \href{mailto:d.j.hessen@uu.nl}{\texttt{d.j.hessen@uu.nl}}
\and Aike N. Vonk\\
   {\small\raggedright Utrecht University}\\
   \href{mailto:a.n.vonk@uu.nl}{\texttt{a.n.vonk@uu.nl}}
\and Peter G. M. van der Heijden\\
    {\small\raggedright Utrecht University and University of Southampton}\\
\href{mailto:p.g.m.vanderheijden@uu.nl}{\texttt{p.g.m.vanderheijden@uu.nl}}
    }
    
\date{}

\date{\vspace{1ex}}

\begin{document}

\maketitle
\begin{abstract}
Correspondence analysis (CA) is a popular technique to visualize the relationship between two categorical variables. CA uses the data from a two-way contingency table and is affected by the presence of outliers. The supplementary points method is a popular method to handle outliers. Its disadvantage is that the information from entire rows or columns is removed. However, outliers can be caused by  cells only. In this paper, a reconstitution algorithm is introduced to cope with such cells. This algorithm can reduce the contribution of cells in CA instead of deleting entire rows or columns. Thus the remaining information in the row and column involved can be used in the analysis. The reconstitution algorithm is compared with two alternative methods for handling outliers, the supplementary points method and MacroPCA. It is shown that the proposed strategy works well.
\end{abstract}

\section{Introduction}

Correspondence analysis (CA) is an exploratory data analysis method which visualizes the dependence of the two categorical variables in a two-way contingency table using a two-dimensional plot \citep{greenacre1984theory, Greenacre1987, greenacre2017correspondence}. CA has received considerable attention in a variety of areas such as marketing \citep{pitt2020new}, psychology \citep{kim2021age}, and text categorization and authorship attribution \citep{Qi_Hessen_Deoskar_vanderHeijden_2023}.
However, relatively little attention has been given to CA in the presence of outliers \citep{Rianietal2022}.

Outliers may be errors or unexpected observations which could shed new light on the researched phenomenon \citep{sripriya2018detection}. In general, the data are arranged in a matrix where rows correspond to the individual observations and columns are variables  \citep{grubbs1969procedures, rousseeuw2018detecting, hubert2019macropca, raymaekers2023challenges}. The term outlier typically refers to a individual observation that deviates markedly from other members of the sample in which it occurs. %Outliers may be errors or accurate but unexpected observations which could shed new light on the researched phenomenon \citep{achtert2010visual, sripriya2018detection}.

In a contingency table, the definition of an outlier is different from the general case \citep{kuhnt2014outlier, sripriya2018detection}. An entry in the table represents the number of individuals that occurs jointly in a category of one variable and a category of the other. Thus, in the contingency table, a row does not correspond to a single observation but to a number of joint sample frequencies of individual observations. Here, extreme counts that do not follow the general pattern in the table are viewed as outliers.

In the context of CA, an outlier can be defined in different ways and the procedure to detect outliers depends on the definition of an outlier.  Two detection procedures stand out. On the one hand, \citet{greenacre2013contributions, greenacre2017correspondence} uses visual inspecting of CA plots to detect outliers. \citet{greenacre2013contributions, greenacre2017correspondence} considers a row or column point as an outlier when it clearly lies far from other points in the CA plot. In addition to  large absolute coordinates, \citet{Donna1986correspondence} and \citet{ bendixen1996practical}  define a row or column point as an outlier if the row or column point has a high contribution to an axis. The contribution of a point to an axis is determined not only by the position of the point in the CA plot but also by the marginal proportion of the point. According to \citet{Donna1986correspondence} and \citet{ bendixen1996practical}, if the marginal proportion of a point is very small, it may not be an outlier, even though, following Greenacre's definition, it is an outlier in the sense that it lies far from other points in the CA plot.

On the other hand, \citet{Rianietal2022} and \citet{ raymaekers2023challenges} detect outliers making use of distributional assumptions. \citet{Rianietal2022} state (p. 8) "... an outlier is a row which does not agree with the multiplicative model assuming independence fitted to the data."  This outlier detection procedure is less attractive, because, in interesting applications, the independence model assumption would be rejected almost always \citep{de1990latent}, and thus, in this situation, this procedure tends to detect too many rows as outlying points. \citet{raymaekers2023challenges} use MacroPCA to detect outliers. MacroPCA is originally proposed by \citet{rousseeuw2018detecting} for principal component analysis (PCA) and subsequently used in CA by \citet{raymaekers2023challenges}. MacroPCA assumes that the data are generated from a multivariate Gaussian distribution. However, the two variables in the contingency table are categorical variables, and therefore the normality assumption for the input matrix of MacroPCA may be not appropriate for CA.

\citet{Donna1986correspondence}, \citet{bendixen1996practical}, \citet{greenacre2017correspondence} and \citet{Rianietal2022} detect outlying rows or columns, and, after detecting the outliers, they cope with the outliers by the supplementary points method. That is, CA is performed on the contingency table without the outlying rows and columns. Afterwards, the outliers are projected into the   CA solution of the reduced table. Therefore, the outliers cannot determine the CA solution. 

In contrast, \citet{raymaekers2023challenges} detect outlying cells and outlying rows and  handle the outliers in the same step. 
The basic idea is to impute the outlying cells by an iterative PCA algorithm while excluding outlying rows.
Their method does not have a good fit with the theory of CA, and  important properties of CA, such as that Euclidean distances in a CA display can be interpreted in terms of chi-squared distances, are lost. Moreover, this method seems to flag a lot more rows as outliers than necessary. 

The supplementary points method and MacroPCA delete outlying rows or columns completely, and therefore, also remove information from these rows or columns that is not related to this outlying problem. So, the removal of an entire row or column causes a unnecessary loss of information. 

According to \citet{bendixen1996practical}, a cell frequency that causes its row to be identified as an outlier might also cause its column to be identified as an outlier, and vice versa. Thus, an outlying row or column may be caused by a specific joint frequency. This suggests that we only need to deal with the specific cell and do not need to delete the entire row or column. 

In this paper, the focus is on cell-wise outliers. To detect outlying cells, we follow Greenacre's definition using visual inspection of the CA plot, as a main aim of CA is to summarize the structure of data via a two-dimensional plot and such outliers cause the other points to be tightly clustered and thus reduce the readability of a CA plot.
A cell is an outlying cell if the corresponding row and column points of this cell lie far from other points. Here, once a cell is identified as an outlier, the cell is not removed but its contribution is reduced. For reducing the contribution of an outlying cell the reconstitution algorithm is proposed. The reconstitution algorithm has been proposed originally by \citet{nora1974methode} and has later been used by \citet{greenacre1984theory} and \citet{de1988correspondence} to handle missing values in cells. 

The paper is built up as follows. We start with a description of CA in Section~\ref{S: CA}. Section~\ref{S: howoutori} presents how outliers originate. Section~\ref{S: rech} presents the reconstitution algorithm to handle cell-wise outliers and describes MacroPCA and the supplementary points method. Section~\ref{S: empstu} compares these three methods on a contingency table, the brands of cars dataset, and compares the reconstitution algorithm and the supplementary points method on an incidence table, the ocean plastic dataset.
%Section~\ref{S: rimp} describes interactive implement. 
Section~\ref{S: con} discusses and concludes this paper. Finally, Section~\ref{S: sof} introduces the implementation of code.

\section{Correspondence analysis background}\label{S: CA}

Let $\bm{X}$ be a contingency table having $I$ rows and $J$ columns with non-negative entries $x_{ij}$, and suppose that $\bm{X}$ has full rank. An index is replaced by '+' when
summed over the corresponding elements, such as $x_{i+} = \sum_jx_{ij}$. It is customary to rescale $\bm{X}$ to the correspondence matrix $\bm{P} = \bm{X}/x_{++}$, so that $\sum_{i}\sum_{j}p_{ij} = 1$. The row profile for row $i$ is the vector having elements $p_{ij} / p_{i+}, j = 1, \dots, J$ and, similarly, the column profile for column $j$ is the vector with elements $p_{ij} / p_{+j}, i = 1, \dots, I$. The average row profile is the vector with elements $p_{+j}, j = 1, \dots, J$, i.e., the column margins, and the average column profile is the vector with elements $p_{i+}, i = 1, \dots, I$, i.e. row margins. Let $\bm{E} = [p_{i+}p_{+j}]$ be the matrix with elements under statistical independence. Let $\bm{D}_r$ and $\bm{D}_c$ be diagonal matrices with the row margins $p_{i+}$ and column margins $p_{+j}$ in the diagonal, respectively. 

CA can be introduced in many ways. We introduce CA here using the concept of total inertia \citep{greenacre2017correspondence}, i.e., the well-known Pearson $\chi^2$ statistic divided by $x_{++}$ %. Total inertia is defined as
    \begin{equation}
   \label{Eq: inertia1}
    \text{Total inertia} = \sum_i\sum_j\frac{(p_{ij}-p_{i+}p_{+j})^2}{p_{i+}p_{+j}}.
  \end{equation}
The aim of CA is to provide a multidimensional representation of the matrix $\bm{X}$ where the total inertia is projected as much as possible onto a low-dimensional space.
The computational procedure to obtain the solution makes use of the singular value decomposition (SVD). In the first step the matrix $\bm{X}$ is transformed into the matrix of standardized residuals $\bm{D}_r^{-\frac{1}{2}}(\bm{P}-\bm{E})\bm{D}_{c}^{-\frac{1}{2}}$ with elements $(p_{ij} - p_{i+}p_{+j}) / \sqrt{p_{i+}p_{+j}}$, and 
then SVD is applied to this matrix, yielding
\begin{equation}
\label{Eq: CA2}
\bm{D}_r^{-\frac{1}{2}}(\bm{P}-\bm{E})\bm{D}_{c}^{-\frac{1}{2}} = \bm{U}\bm{\Sigma}\bm{V}^T, 
\end{equation}
where $\bm{U}^T\bm{U} = \bm{V}^T\bm{V} = \bm{I}$ and $\bm{\Sigma}$ is a diagonal matrix with singular values $\sigma_k, k = 1, \cdots, \text{min}(I-1, J-1)$ in descending order  on the diagonal. Subtracting the matrix $\bm{E}$ of rank 1 leads to a reduction of 1 for the rank of the resulting matrix $\bm{D}_r^{-\frac{1}{2}}(\bm{P}-\bm{E})\bm{D}_{c}^{-\frac{1}{2}}$.

If we pre-multiply and post-multiply both sides of Equation~(\ref{Eq: CA2})
by $\bm{D}_r^{-\frac{1}{2}}$ and $\bm{D}_c^{-\frac{1}{2}}$, respectively, on the left hand side we get $\bm{D}_r^{-1}(\bm{P}-\bm{E})\bm{D}_{c}^{-1}$ with elements $(p_{ij} - p_{i+}p_{+j}) / {p_{i+}p_{+j}}$, and this yields
\begin{equation}
\label{Eq: CA3}
\bm{D}_r^{-1}(\bm{P}-\bm{E})\bm{D}_{c}^{-1} = \bm{D}_r^{-\frac{1}{2}}\bm{U}\bm{\Sigma}(\bm{D}_{c}^{-\frac{1}{2}}\bm{V})^T = \bm{\Phi}\bm{\Sigma}\bm{\Gamma}^T = \bm{F}\bm{\Sigma}^{-1}\bm{G}^T
\end{equation}
where $\bm{\Phi} = \bm{D}_r^{-\frac{1}{2}}\bm{U}$, $\bm{\Gamma} = \bm{D}_c^{-\frac{1}{2}}\bm{V}$, $\bm{F} = \bm{\Phi}\bm{\Sigma}$, and $\bm{G} = \bm{\Gamma}\bm{\Sigma}$. $\bm{\Phi}$ and $\bm{\Gamma}$ are called the standard coordinates for the row profiles and column profiles, respectively. They have the property that, for each $k$, their weighted sum is 0 and their weighted sum of squares is 1, i.e. $
 \bm{1}^T\bm{D}_r\bm{\Phi} = \bm{1}^T\bm{D}_c\bm{\Gamma} = \bm{0}^T$ and $\bm{\Phi}^T\bm{D}_r\bm{\Phi} = \bm{\Gamma}^T\bm{D}_c\bm{\Gamma} = \bm{I}$. 
 $\bm{F}$ and $\bm{G}$ are called principal coordinates for the row profiles and column profiles, respectively.

Euclidean distances between rows of $\bm{F}$ ($\bm{G}$) are equal to the so-called $\chi^2-$ distances between rows (columns) of $\bm{X}$. 
The squared $\chi^2-$distance between the row profiles $i$ and $i'$ is
\begin{equation}
    \delta^2_{i,i'} = \sum_j\frac{(\frac{p_{ij}}{p_{i+}} - \frac{p_{i'j}}{p_{i'+}})^2} {p_{+j}}.
\end{equation}
The $\chi^2-$distance $\delta_{i,i'}$ between row profiles $i$ and $i'$ gives more weight to differences in a column $j$ when this column has a lower margin $p_{+j}$. The $\chi^2-$distance $\delta_{j,j'}$ between column profiles $j$ and $j'$ is defined in a similar way.

Joint graphic displays of row points and column points are usually made to study the relationship between the rows and the columns in the matrix $\bm{P}$. For this asymmetric and symmetric maps are used. In an asymmetric map rows of $\bm{P}$ can be displayed as points in a multidimensional space using principle coordinates, and columns as points using standard coordinates. Thus, in full-dimensional space the dot products of row points $\bm{F}$ and column points $\bm{\Gamma}$ are equal to the elements of $\bm{D}_r^{-1}(\bm{P}-\bm{E})\bm{D}_{c}^{-1}$. Usually low-dimensional representations are made of the first few columns of $\bm{F}$ and  $\bm{\Gamma}$, as the SVD ensures that the first few dimensions provide an optimal approximation of $\bm{D}_r^{-1/2}(\bm{P}-\bm{E})\bm{D}_{c}^{-1/2}$ in a least-squares sense. Together, the configurations of row points and column points form a biplot \citep{GabrielBiplot1971} of the matrix $\bm{D}_r^{-1}(\bm{P}-\bm{E})\bm{D}_{c}^{-1}$. Asymmetric maps have the interesting property that the row points are in the weighted average of the column points and the other way around. This is evident from the so-called transition equations

\begin{equation}
\label{Eq: CA5}
\begin{split}
\bm{F} = \bm{D}_r^{-1}\bm{P}\bm{\Gamma}~~~~~~ \text{and}~~~~~~\bm{G} = \bm{D}_c^{-1}\bm{P}^T\bm{\Phi}
\end{split}
\end{equation}

The points for the average row profile and the average column profile  fall in the origin. Thus, for the combination of $\bm{F}$ and  $\bm{\Gamma}$, the transition formulas pull individual row points towards the column points for which $p_{ij}/p_{i+} > p_{+j}$.

Asymmetric maps have the drawback that, when for example the pair $\bm{F}$ and  $\bm{\Gamma}$ is used, the Euclidean distances between the columns are not chi-squared distances. Also, there is the practical disadvantage that the cloud of column points may be huge in comparison to the cloud of row points, and thus row points tend to huddle together and reduce the readability of the plot. For this reason one often sees the use of the so-called symmetric map. That is, both rows and columns are displayed in principle coordinates. Therefore, the Euclidean distances between row points, i.e., rows of $\bm{F}$ (column points, i.e., rows of $\bm{G}$) are equal to the $\chi^2-$distances between rows (columns) of $\bm{X}$, and in low-dimensional representations the Euclidean distances between row points and between column points provide approximations of these distances. The Euclidean distance between row points and column points is not meaningful. However, the direction between row points and column points is still meaningful, because the only difference between principal and standard coordinates is a dimensionwise scalar (compare Equation~(\ref{Eq: CA3})).

The total inertia can be expressed as a weighted sum of squared $\chi^2-$distances of row profiles and of column profiles to the average profile:
\begin{equation}
\label{Eq: inertiachi}
\text{Total inertia} = \sum_ip_{i+}\sum_j\frac{(\frac{p_{ij}}{p_{i+}} - p_{+j})^2}{p_{+j}} = \sum_jp_{+j}\sum_i\frac{(\frac{p_{ij}}{p_{+j}} - p_{i+})^2}{p_{i+}}.
\end{equation}
This shows that the total inertia can be split up over the rows 
and over the columns. The inertia of the row point $i$ and the column point $j$ in dimension $k$ are $p_{i+}f_{ik}^2 = u_{ik}^2\sigma_k^2$ and $p_{+j}g_{jk}^2  = v_{jk}^2\sigma_k^2$, respectively. The contributions of row $i$ and column $j$ to dimension $k$ are  $p_{i+}f_{ik}^2 / \sigma_k^2 = u_{ik}^2\sigma_k^2 / \sigma_k^2 = u_{ik}^2$ and $p_{+j}g_{jk}^2 / \sigma_k^2 = (v_{jk}\sigma_k)^2 / \sigma_k^2 = v_{jk}^2$, respectively. The contributions quantify to what extent individual rows and columns, both by their positions ($f_{ik}$ or $g_{jk}$) and margins ($p_{i+}$ or $p_{+j}$), affect the solution \citep{greenacre2013contributions}. This means that, for rows that have equal margins $p_{i+}$ for dimension $k$, the further this point is from the origin, the larger  its contribution is to dimension $k$. 
In a so-called \emph{contribution biplot}, elements $f_{ik}$ ($u_{ik}$) are as row coordinates and $v_{jk}$ ($g_{jk}$) as column coordinates. 

The total inertia can also be split up over cells. The inertia of each cell in the matrix $\bm{D}_r^{-\frac{1}{2}}(\bm{P}-\bm{E})\bm{D}_{c}^{-\frac{1}{2}}$ of standardized residuals is $(p_{ij}-p_{i+}p_{+j})^2 / p_{i+}p_{+j}$.

By rewriting Equation~(\ref{Eq: CA3}), the correspondence matrix $\bm{P}$ can be decomposed as follows
\begin{equation}
\label{Eq: CA6}
\bm{P} = \bm{D}_r(\bm{1}\bm{1}^T + \bm{\Phi}\bm{\Sigma}\bm{\Gamma}^T)\bm{D}_{c} \approx \bm{D}_r(\bm{1}\bm{1}^T + \bm{\Phi}_K\bm{\Sigma}_K\bm{\Gamma}_K^T)\bm{D}_{c}
\end{equation}
Equation~(\ref{Eq: CA6}) is called the reconstitution formula and is the foundation of the \emph{reconstitution algorithm}, discussed in Section~\ref{S: rech}.

Similar to Equation~(\ref{Eq: CA5}), an additional row can be projected as a supplementary point in an existing CA plot. Let the extra row (supplementary) point be the row vector $\bm{a} = [a_1, a_2, \cdots, a_J]$ and an extra column (supplementary) point $\bm{b} = [b_1, b_2, \cdots, b_I]$, as a row vector. The projections for the row point $\bm{a}$ and the column point $\bm{b}$ are found by
\begin{equation}
\label{Eq: CAproject}
\frac{\bm{a}}{\sum_{j}a_j}\bm{\Gamma}~~~~~~\text{and} ~~~~~~
\frac{\bm{b}}{\sum_ib_i}\bm{\Phi}
\end{equation}
respectively. These supplementary points do not determine the CA solution, but from these projections we can see the relationships between the configurations of row and column points in the existing CA solution to these supplementary points.

\section{How outliers originate}\label{S: howoutori}

CA is sensitive to outliers \citep{Choulakian_2020}. Here, we enumerate three potential causes for the presence of outliers: an (approximate) block diagonal matrix, rows or columns with relatively small margins, and cells with relatively high values. We do not claim that these causes give an exhaustive view, and also note that these three causes may overlap.

\subsection{Block diagonal matrix}

As we discussed in Equation~(\ref{Eq: inertiachi}), the total inertia can be expressed as a weighted sum of squared $\chi^2-$distances of rows points to the origin \citep{greenacre2017correspondence}. If all profiles are the same and thus equal to the average profile, then all $\chi^2$-distances of the points to the origin would be 0 and thus the total inertia would be 0. On the other hand, maximum inertia can be obtained when all profiles are totally different. For example, when a matrix is an identity matrix and $m = n$, the inertia is equal to $m-1$. 

If, after reordering the rows and columns of a matrix in an appropriate way, $\bm{X}$ is a block diagonal matrix with $t$ blocks, the first $t - 1$ dimensions of the CA solution have singular values equal to $1$ \citep{Choulakian_2020, sander2023}. For example, let $t = 2$. Table~\ref{madeupblock} is an illustration. One block is cell (2, b) and the other block consists of rows 1, 3 and 4 together with columns a, c, d.
The CA solution is as follows

\begin{equation}
\begin{aligned}
\bm{D}_r^{-1}(\bm{P}-\bm{E})\bm{D}_{c}^{-1}
&= \bm{\Phi}\bm{\Sigma}(\bm{\Gamma})^T  \\&= 
\left[
 \begin{array}{rrrrrr}
0.27 & 1.59 & 0.18 \\ 
  -3.67 & 0.00 & 0.00 \\ 
  0.27 & -0.50 & -1.52 \\ 
  0.27 & -0.79 & 0.97 \\ 
  \end{array}
  \right]\left[
 \begin{matrix}
 1  & 0 & 0\\
 0 & 0.44 & 0 \\
 0 & 0 & 0.10 \\
  \end{matrix}
  \right] 
 \left[
 \begin{array}{rrrrrr}
0.27 & -0.83 & 1.54 \\ 
  -3.67 & 0.00 & 0.00 \\ 
  0.27 & -0.40 & -0.91 \\ 
  0.27 & 1.91 & 0.34 \\ 
  \end{array}
  \right]^T  
\end{aligned}
\end{equation}
The first singular value equals 1. 
The reordered matrix based on the first coordinates of the rows and columns, shown in Table~\ref{madeupblockreorder}, is a block diagonal matrix. On dimension 1 row 2 and column 2 are outliers with scores -3.67.

Table~(\ref{madeupapproblock}) is a less extreme case, where the elements approximate a block diagonal matrix. The CA solution is as follows

\begin{equation}
\begin{aligned}
\bm{D}_r^{-1}(\bm{P}-\bm{E})\bm{D}_{c}^{-1}
&= \bm{\Phi}\bm{\Sigma}(\bm{\Gamma})^T  \\&= 
\left[
 \begin{array}{rrrrrr}
0.19 & 1.56 & 0.18 \\ 
  -5.03 & -0.03 & 0.02 \\ 
  0.20 & -0.49 & -1.49 \\ 
  0.21 & -0.78 & 0.95 \\ 
  \end{array}
  \right]\left[
 \begin{matrix}
0.93  & 0 & 0\\
 0 & 0.44 & 0 \\
 0 & 0 & 0.09 \\
  \end{matrix}
  \right] 
 \left[
 \begin{array}{rrrrrr}
0.20 & -0.82 & 1.52 \\ 
  -5.03 & -0.07 & -0.03 \\ 
  0.21 & -0.39 & -0.90 \\ 
  0.17 & 1.88 & 0.33 \\  
  \end{array}
  \right]^T  
\end{aligned}
\end{equation}
Now the first singular value is 0.93, close to 1. The reordered matrix, where the row and column scores for dimension 1 are used, is in Table \ref{madeupapproblockreorder}. It approximates a block diagonal matrix. On dimension 1 row 2 and column b are outliers with scores -5.03.

By these examples we want to illustrate that, if the rows and columns of the table can be reordered so that a block diagonal matrix or an approximate block diagonal matrix arises, this may lead to outlying points for those rows and columns that form the smaller (approximate) block diagonal matrix.

\begin{table}[h]
\caption{Document-term matrix $\bm{X}$: size 4$\times$4}
    \begin{subtable}[h]{0.45\textwidth}
       \centering  
\begin{tabular}{cccccccc}    
\hline
 & a & b & c & d \\  
\hline  
1 & 1 & 0 & 3 & 4 \\ 
2 & 0 & 2 & 0 & 0\\ 
3 & 2 & 0 & 5 & 1 \\ 
4 & 4 & 0 & 6 & 1 \\ 
\hline 
\end{tabular} 
\caption{Block diagonal matrix} 
\label{madeupblock}
    \end{subtable}
    %\vspace{5mm}
        \begin{subtable}[h]{0.45\textwidth}
       \centering  
\begin{tabular}{cccccccc}    
\hline
 &b &d  & c & a \\  
\hline  
2&2 & 0 & 0 & 0 \\ 
1&  0 & 4 & 3 & 1 \\ 
3&  0 & 1 & 5 & 2 \\ 
4&  0 & 1 & 6 & 4 \\
\hline 
\end{tabular} 
\caption{Block diagonal matrix; reordered table} 
\label{madeupblockreorder}
    \end{subtable}
    \begin{subtable}[h]{0.45\textwidth}
\centering  
\begin{tabular}{ccccccc}    
\hline
 & a & b & c & d\\  
\hline  
1 & 100 & 2 & 300 & 400 \\ 
2 &  2 & 100 & 1 & 4 \\ 
3 &  200 & 3 & 500 & 100 \\ 
4 &  400 & 2 & 600 & 100 \\ 
\hline 
\end{tabular} 
\caption{\footnotesize{Approximate block diagonal matrix}}
\label{madeupapproblock}
     \end{subtable}
    %\vspace{5mm}
        \begin{subtable}[h]{0.45\textwidth}
\centering  
\begin{tabular}{ccccccc}    
\hline
 & b & d & a & c\\  
\hline  
2&100 & 4 & 2 & 1 \\ 
1&  2 & 400 & 100 & 300 \\ 
3&  3 & 100 & 200 & 500 \\ 
4&  2 & 100 & 400 & 600 \\  
\hline 
\end{tabular} 
\caption{\footnotesize{Approximate block diagonal matrix; reordered table}}
\label{madeupapproblockreorder}
     \end{subtable}
    \hfill
    \begin{subtable}[h]{0.45\textwidth}
\centering  
\begin{tabular}{ccccccc}    
\hline
 & a & b & c & d\\  
\hline  
1 &1 & 2 & 3 & 4 \\ 
2&  2 & 100 & 1 & 4 \\ 
3& 2 & 3 & 5 & 1 \\ 
4& 4 & 2 & 6 & 1 \\ 
\hline 
\end{tabular} 
\caption{Large value 100} 
\label{madeuplargevalue}
    \end{subtable}
        \hfill
    \begin{subtable}[h]{0.45\textwidth}
\centering  
\begin{tabular}{ccccccc}    
\hline
 & c & a & d & b\\  
\hline  
4 &6 & 4 & 1 & 2 \\ 
3&  5 & 2 & 1 & 3 \\ 
 1& 3 & 1 & 4 & 2 \\ 
2&  1 & 2 & 4 & 100 \\
\hline 
\end{tabular} 
\caption{Large value 100; reordered table} 
\label{madeuplargevaluereorder}
    \end{subtable}
\label{T: datasets}
\end{table}

\subsection{Rows or columns with relatively small margins}

\noindent For the rows of the matrix $\bm{X}$, 
Equation~(\ref{Eq: inertiachi}), the squared Euclidean distance of row $i$ to the origin $O$, $\delta^2_{iO}$, is 
\begin{equation}
\label{Eq: chi2iO}
\delta^2_{iO} = \sum_j \frac{\left(\frac{p_{ij}}{p_{i+}} - p_{+j}\right)^2}{p_{+j}}.
\end{equation}
Following \citet{greenacre2013contributions}, we will argue that, in principle, rows (and columns) with smaller margins $p_{i+}$ have relatively more potential than rows (and columns) with larger margins to be in the periphery of a cloud of points, as these rows with smaller margins $p_{i+}$ have a higher  potential to have a larger chi-squared distance $\delta^2_{iO}$.
\begin{itemize}
    \item{The marginal profile with elements $p_{+j}$ falls in the origin. Outliers fall relatively far away from the origin. Thus we are interested in what makes the distance of row $i$ from the origin, $\delta^2_{iO}$, larger.}
    \item{In Equation~(\ref{Eq: chi2iO}) $\left(p_{ij}/p_{i+} - p_{+j}\right)^2$ stands for the squared difference between the row profile element $j$ and the marginal profile element $j$.}
    \item{The marginal profile is the weighted average of the row profiles with weights $p_{i+}$, as for each element $j$, $\sum_i p_{i+} (p_{ij}/p_{i+}) = p_{+j}$. }
    \item{Therefore, if row $i$ has a larger size $p_{i+}$, we expect that, in principle, row $i$ will be closer to the marginal profile, as it makes up a larger part of the marginal profile. In other words, row profiles with larger $p_{i+}$ have a larger expected correlation with the marginal profile. {\em So in principle rows with smaller $p_{i+}$ have a higher potential to be relatively further away from the origin.}}
    \item{Now consider the denominator $p_{+j}$ in the squared chi-squared distance. For a fixed difference $\left(p_{ij}/p_{i+} - p_{+j}\right)^2$ columns with smaller $p_{+j}$ add more to the chi-squared distance of row $i$ to the origin.  }
    \item Now consider the above reasoning for the squared chi-squared distance between column $j$ and the origin, $\delta^2_{jO}$. The same argument holds, {\em columns with smaller $p_{+j}$ have in principle a higher potential to be relatively further away from the origin.}
    \item{At the same time we noticed that, for a fixed difference $\left(p_{ij}/p_{+j} - p_{i+}\right)^2$, rows with smaller $p_{i+}$ add more to the chi-squared distance of column $j$ to the origin.}
\end{itemize}

We conclude that smaller margins have the potential to lead to larger chi-squared distances of individual rows and columns to the origin because of their potential to deviate more from the marginal profile falling in the origin, and due to the role of the marginal probability in the denominator. If row $i$ has a large difference 
$\left(p_{ij}/p_{i+} - p_{+j}\right)^2$ in element $j$, and in particular in element $j$ where $p_{+j}$ is small, then it is more likely that an outlier arises. 

We also note that, if row $i$ is an outlier due to profile element $j$ having a low $p_{+j}$, then if profile element $i$ has a low $p_{i+}$, column $j$ will also be an outlier. The reason is that we can formulate independence in three ways, namely as $\left(p_{ij} = p_{i+}p_{+j}\right)$, as
$\left(p_{ij}/p_{i+} = p_{+j}\right)$ and as $\left(p_{ij}/p_{+j} = p_{i+}\right)$. If there is positive dependence, then $\left(p_{ij} > p_{i+}p_{+j}\right)$, then also 
$\left(p_{ij}/p_{i+} > p_{+j}\right)$ and  $\left(p_{ij}/p_{+j} > p_{i+}\right)$. The latter two conditional formulations of positive dependence link directly to the  squared chi-squared distances $\delta^2_{iO}$ to $\delta^2_{jO}$. Thus a single cell $(i, j)$ with a strong positive relation can cause a row $i$ as well as column $j$ to be an outlier.

\subsection{Cells with relatively high values}

Outliers may occur due to relatively large frequencies \citep{Langovayaetal2013, Choulakian_2020}. Table~(\ref{madeuplargevalue}) is an illustration where row 2 and column b have a relative large frequency of 100. The CA solution is
\begin{equation}
\begin{aligned}
\bm{D}_r^{-1}(\bm{P}-\bm{E})\bm{D}_{c}^{-1}
&= \bm{\Phi}\bm{\Sigma}(\bm{\Gamma})^T  \\&= 
\left[
 \begin{array}{rrrrrr}
-1.49 & -3.28 & 0.36 \\ 
  0.56 & 0.05 & 0.04 \\ 
  -1.67 & 0.74 & -2.91 \\ 
  -2.05 & 1.44 & 1.89 \\  
  \end{array}
  \right]\left[
 \begin{matrix}
0.75  & 0 & 0\\
 0 & 0.31 & 0 \\
 0 & 0 & 0.08 \\
  \end{matrix}
  \right] 
 \left[
 \begin{array}{rrrrrr}
-1.76 & 1.44 & 3.08 \\ 
  0.55 & 0.12 & -0.07 \\ 
  -2.18 & 0.55 & -1.83 \\ 
  -0.99 & -3.41 & 0.71 \\ 
  \end{array}
  \right]^T  
\end{aligned}
\end{equation}
The reordered matrix based on the first coordinates of the rows and columns is shown in Table~\ref{madeuplargevaluereorder}. On dimension 1 row 2 and column b are outliers with scores 0.56 and 0.55, respectively. 

\section{Methods to handle outliers}\label{S: rech}

We discuss three methods to handle outliers. Two methods are cell-wise outlier methods: reconstitution of order $h$ and MacroPCA. The third is the supplementary points method. It is worth noting that reconstitution of order $h$ has been used to handle missing data, but has not been proposed to handle outliers.

\subsection{Reconstitution of order $h$}

In this paper we propose to deal with an outlier or outliers by changing the data. Specifically, we assume that specific cells in a matrix are outlying cells if they cause row and column points to be outliers. We propose to make such cells in the data matrix missing. We use visual inspection of the CA plot to define outlying cells. In a second step, we apply an algorithm that imputes a new value for each missing value. For this, we use the reconstitution algorithm, originally proposed by \citet{nora1974methode} and revisited by \citet{greenacre1984theory}, \citet{de1988correspondence}, and \citet{josse2012handlingmca}.

We assume for the moment that there is only a single cell causing a row and a column to be outliers, but the procedure that we describe can be applied to multiple outlying cells simultaneously. The idea is to adjust the value in this single cell in such a way that it is perfectly reconstituted in a $h$-dimensional CA solution. This reconstitution is obtained iteratively.

As by iteratively imputing the missing cell the margins also change, it is easier to describe the method using the raw data $x_{ij}$ instead of the proportions $p_{ij}$. For $x_{ij}$ we have 

\begin{equation}\label{E: xrec}
   x_{ij} =  \frac{x_{i+}x_{+j}}{x_{++}}\left(1 + \sum_{k = 1}^{\text{min}\{I-1, J-1\}}\phi_{ik}\sigma_k\gamma_{jk}\right),
\end{equation}

\noindent i.e. $x_{ij}$ is reconstituted if the maximum dimensionality min $(I-1, J-1)$ is used.
Let $\hat{x}_{ij}$ be the reconstituted value using $h <$  min $(I-1, J-1)$ dimensions. Then
\begin{equation}\label{E: xappro}
   \hat{x}_{ij} =  \frac{x_{i+}x_{+j}}{x_{++}}\left(1 + \sum_{k = 1}^{h}\phi_{ik}\sigma_k\gamma_{jk}\right).
\end{equation}

We first explain reconstitution of order 0, meaning that no CA dimensions are used in the reconstitution. Assume that cell $(m,n)$ is an outlier made missing, and assume that at iteration $t=0$ we impute a non-negative value. Then we iteratively find updates for this missing value as follows: 
\begin{equation}
x_{mn}^{t+1} = \frac{x_{m+}^{t}x_{+n}^{t}}{x_{++}^{t}}.
\end{equation}
After convergence, we have the converged value $x_{mn}^*$. Then CA is applied to the original data where the outlier value in cell $(m,n)$ is replaced by $x_{mn}^*$.
As $x_{mn}^* = x_{m+}^*x_{+n}^*/x_{++}^*$, in (\ref{E: xrec}) the residual for cell $(m,n)$ $x_{mn}^* - x_{m+}^*x_{+n}^*/x_{++}^* = 0$. In this sense, the influence of the original outlying cell is eliminated. \citet{de1988correspondence} use reconstitution of order zero in the context of the statistical quasi-independence model. They adjust CA so that it can decompose the departure from this model, a model that assumes independence for some but not all cells in a contingency table. Reconstitution of order zero is also availabe in the R Package {\em anacor} \citep{de2009simple}. 

However, as the residual for cell $(m,n)$ is 0, the inner-product $\sum_{k = 1}^{\text{min}\{I-1, J-1\}}\phi^*_{mk}\sigma^*_k\gamma^*_{nk} = 0$ as well, meaning that in the full-dimensional space the vectors $m$ and $n$ are orthogonal. This may be an undesirable bi-product of reconstitution of order 0. An alternative, reconstitution of order $h$, does not have this problem. In reconstitution $h$, the value in cell $(m,n)$ is reconstituted by

\begin{equation}\label{E: xappro-h}
   x_{mn}^{t+1} =  \frac{x_{m+}^tx_{+n}^t}{x_{++}^t}\left(1 + \sum_{k = 1}^{h}\phi_{mk}^t\sigma_k^t\gamma_{nk}^t\right).
\end{equation}

\noindent Thus in the $h$-dimensional solution the value in cell $(m,n)$ is reconstituted perfectly by $x^*_{m,n} = \left(x^*_{m+}x^*_{+n}/x^*_{++}\right)(1+\sum_{k = 1}^{h}\phi^*_{mk}\sigma_k^*\gamma_{nk}^*)$, and only for higher dimensions than $h$ the residual as well as inner-product is zero. This means that the parameters $\phi^*_{ik}$, $\sigma^*_k$, and $\gamma^*_{jk}$, $k = 1, 2, \cdots, h$ provide the CA solution based on the non-outlying cells in the matrix only. 
So, when interest goes out to a CA solution of two dimensions, theoretically it makes sense to eliminate the influence of an outlier by applying reconstitution of order 2. However, in practice this may lead to a negative value for $x^*_{m,n}$, as is the case in the second example of Section~\ref{S: empstu}. In such instances reconstitution of order zero is the  preferred option.

As far as we know, there is no R package in which reconstitution of order $h$ is implemented, where $h \geq 1$. We present the R function {\em reconca}, that we created by rewriting the function {\em imputeCA} taken from the R package {\em missMDA} which implements a regularized reconstitution algorithm \citep{josse2012handlingmca, josse2016missmda} that is meant for the missing value problem where the number of missing values in the data is relatively large. This is a situation different from our idea to make outlying values missing and therefore we further ignore this regularized version in this paper.

\subsection{MacroPCA}
 
MacroPCA was originally proposed for PCA \citep{hubert2019macropca} and subsequently adjusted for CA \citep{raymaekers2023challenges}. MacroPCA is quite involved and detects outliers and handles outliers at the same time. It includes two parts. The first part of MacroPCA is a multivariate method called DetectDeviatingCells (DDC) \citep{rousseeuw2018detecting, hubert2019macropca} that assumes that data are generated from a multivariate Gaussian distribution but some cells were corrupted.
DDC detects cellwise outliers, and provides these cellwise outliers with initial values. It also detects initial row-wise outliers. In the second part, the set of outlying rows will be improved. Low-dimensional representations are obtained in a way that is similar but not identical to the reconstitution algorithm. The low-dimensional representations of MacroPCA are not nested. That is, for example, the two-dimensional representation is not a subset of three-dimensional representations. We refer to \citet{rousseeuw2018detecting, hubert2019macropca} for details. 

MacroPCA is modified to handle missing data and outlier problems in the context of CA \citep{raymaekers2023challenges}. For CA the original matrix is replaced with the matrix of standardized residuals. As in CA the standardized residuals are only a starting point in finding the CA solution, the modification is close to but different from CA. Also, in the DCC step of MacroPCA where outlying cells are detected, the
algorithm makes the assumption of a Gaussian distribution, for which there is no clear rationale in the context of CA.

\subsection{Supplementary points method}

The supplementary points method is a well known method to deal with row-wise outliers or column-wise outliers. That is, after noticing outlying points, for which we use visual inspection, a new CA is performed on the data matrix where these row-wise or column-wise outliers are removed. Then, as a second step, these outliers are projected as supplementary points into the existing CA solution. Using Equation~(\ref{Eq: CAproject}) in Section~\ref{S: CA}, if an outlier $\bm{a}$ is a row point, its coordinates in the $K$-dimensional CA solution are given by $(\bm{a} / \sum_{j}a_j)\bm{\Gamma}_K$ and if an outlier $\bm{b}$ is a column point, its coordinates in the $K$-dimensional CA solution are given by $(\bm{b} / \sum_ib_i)\bm{\Phi}_K$. 

The supplementary points method is a standard method to deal with outliers in CA, see, for example, \citet{Donna1986correspondence}, \citet{bendixen1996practical}, \citet{greenacre2017correspondence}, and \citet{Rianietal2022}. However, as we argued above, outliers may be caused by a single cell in the data matrix, and deleting an entire row or column where cell-wise outliers occur from the contingency table leads to a loss of the entire category, including outlying and non-outlying cells. 
In contrast, reconstitution of order $h$ eliminates the effect of only the outlying cells, thus keeping as much information as possible in the analysis.

\section{Empirical studies/Results}\label{S: empstu}

We consider two datasets, the attributes of brands of cars and ocean plastic datasets. The attributes of brands of cars dataset is a classic dataset to study CA with the problem of outliers, see, for example, \citet{Rianietal2022, raymaekers2023challenges}. Therefore, we compare reconstitution of order $h$, MacroPCA, and the supplementary points method on this dataset.

The ocean plastic dataset is an incidence dataset created by \citet{vonk2024comparative}. We use this dataset to show that the reconstitution algorithm is appropriate for incidence data as well. However, we do not discuss MacroPCA for this example, as MacroPCA applied to this dataset yielded a degenerate solution (See Supplementary materials).  The reason for this is not clear to us, but we notice that assumptions underlying MacroPCA are severely violated by the matrix of standardized residuals. Therefore, for this dataset, we only compare reconstitution of order $h$ and the supplementary points method. 

\subsection{The attributes of brands of cars data}

As a first dataset,  we use the attributes of brands of cars dataset to illustrate our method. The dataset has been analysed before in \citet{Rianietal2022, raymaekers2023challenges}. This dataset is a part of the R package {\em cellWise} \citep{Raymaekers2023cellWise}. See Table~\ref{T: Tbrands}  for the data. The contingency table consists of 39 rows and 7 columns. The rows represent 39 brands of cars, such as {\em Jeep}, {\em Porsche}, and {\em Volvo}. The seven columns represent the attributes: {\em Fuel Economy}, {\em 
Innovation}, {\em Performance}, {\em Quality}, {\em Safety}, {\em Style}, and {\em Value}. In total 1,578 participants were asked what they considered
attributes for the 39 different vehicle brands. They selected all attributes in the list which they felt applied to a brand. An entry in the  table represents the number of respondents that chose the attribute for a car. In total this led to 11,713 scorings. We note that this is not a typical contingency table as in a typical table the total count is identical to the number of respondents.

\begin{sidewaysfigure}
\caption{Car dataset CA plot}\label{F: cabrandscasym}
\centering
    \begin{subfigure}[b]{0.35\linewidth}
        \includegraphics[width=1\textwidth]{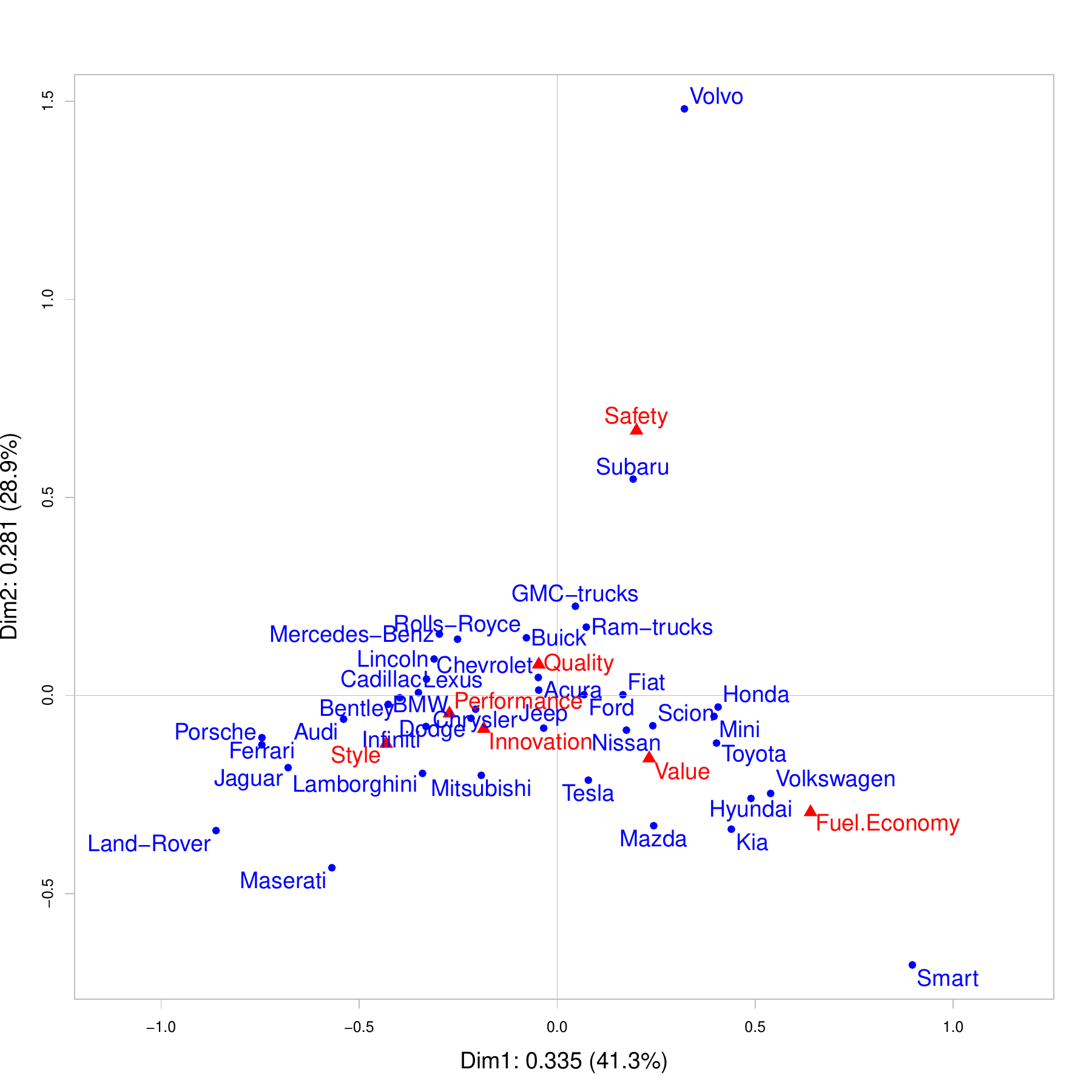}
         \caption{CA original table}\label{F: brandscasym}
    \end{subfigure}
       \begin{subfigure}[b]{0.35\linewidth}
             \includegraphics[width=1\textwidth]{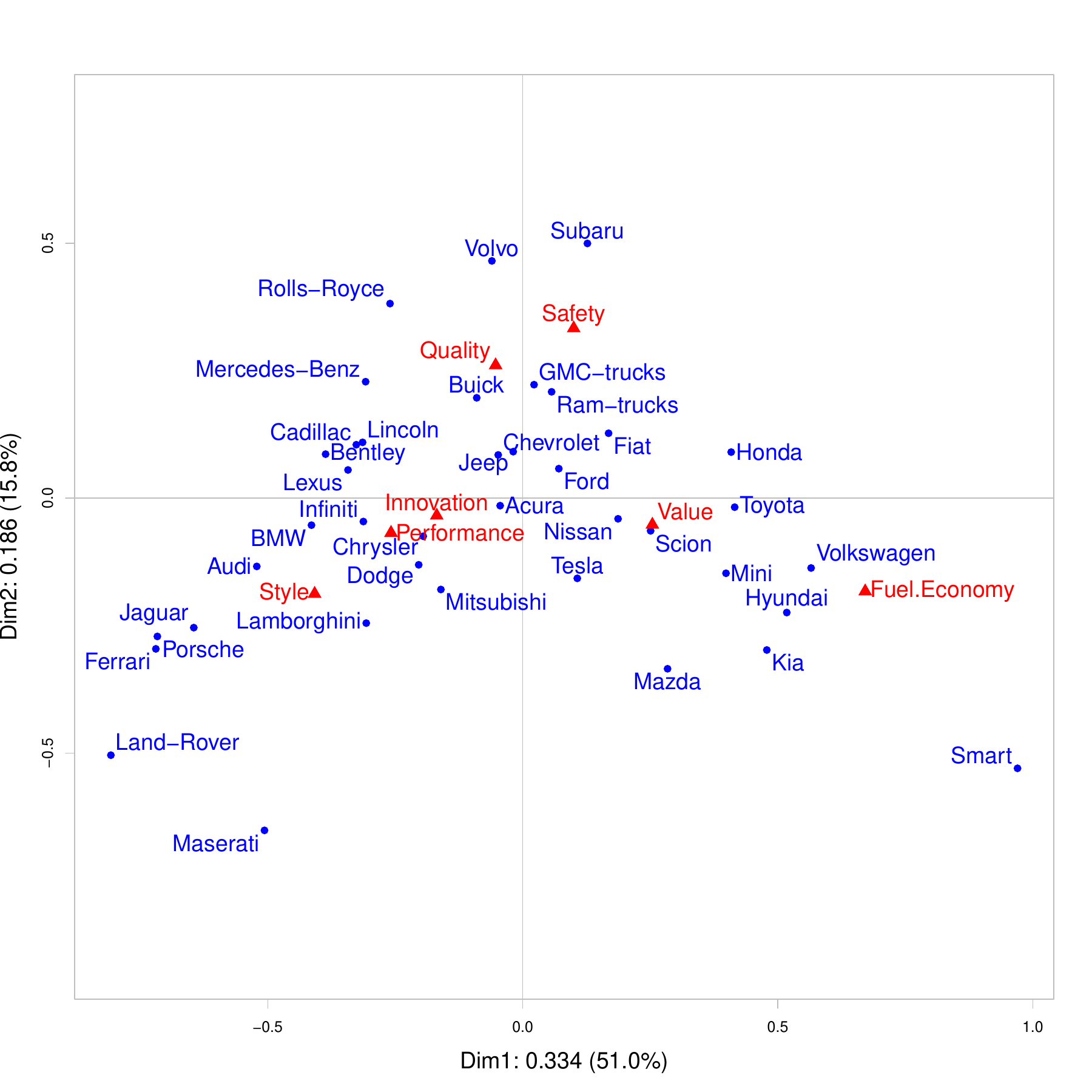}
         \caption{\scriptsize CA reconstitution of order 2 (Volvo, safety)}\label{F: recbrandscasym}
    \end{subfigure}
        \begin{subfigure}[b]{0.35\linewidth}
            \includegraphics[width=1\textwidth]{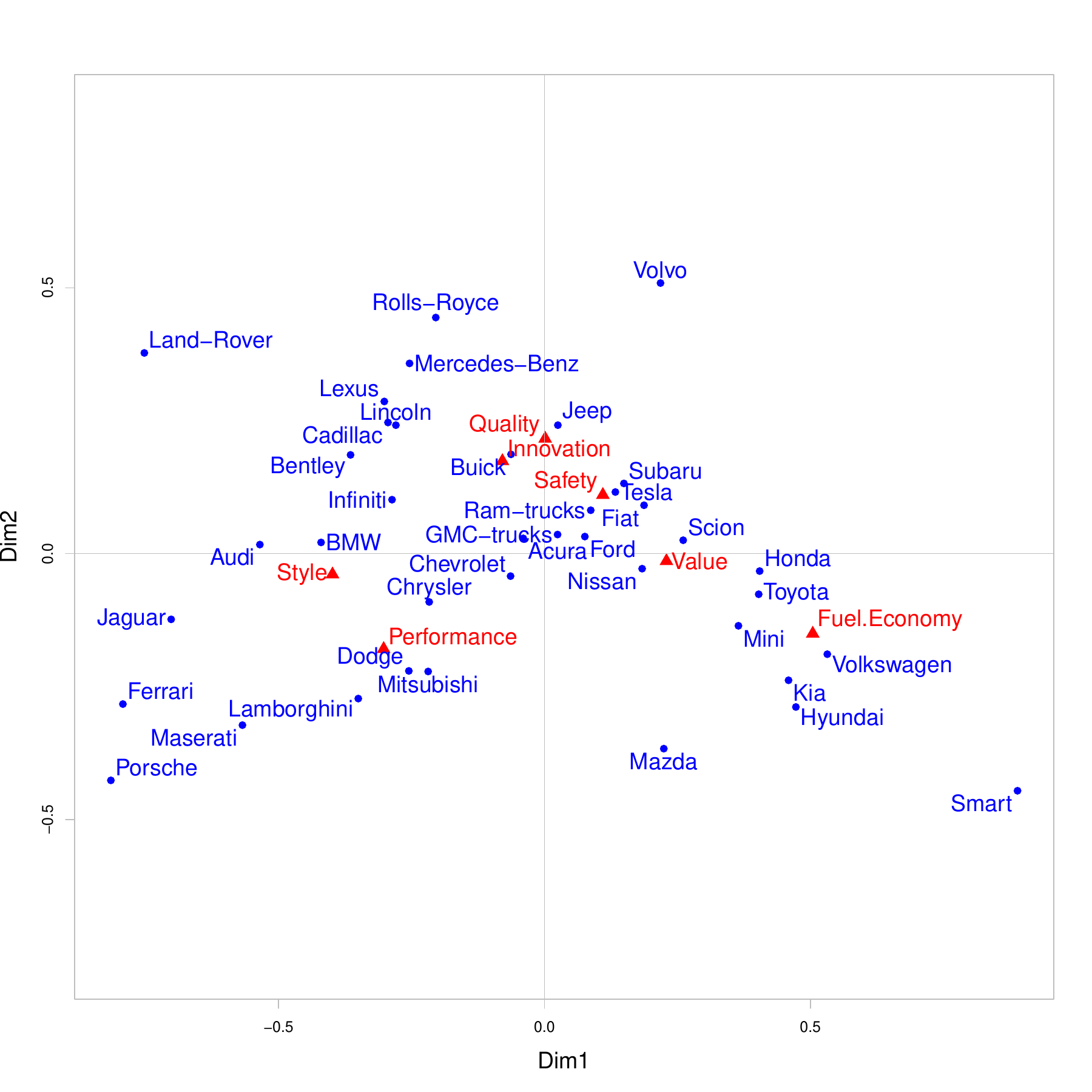}
    \caption{CA MacroPCA of order 2}\label{F: macbrandscasym}
    \end{subfigure}
    \begin{subfigure}[b]{0.35\linewidth}
          \includegraphics[width=1\textwidth]{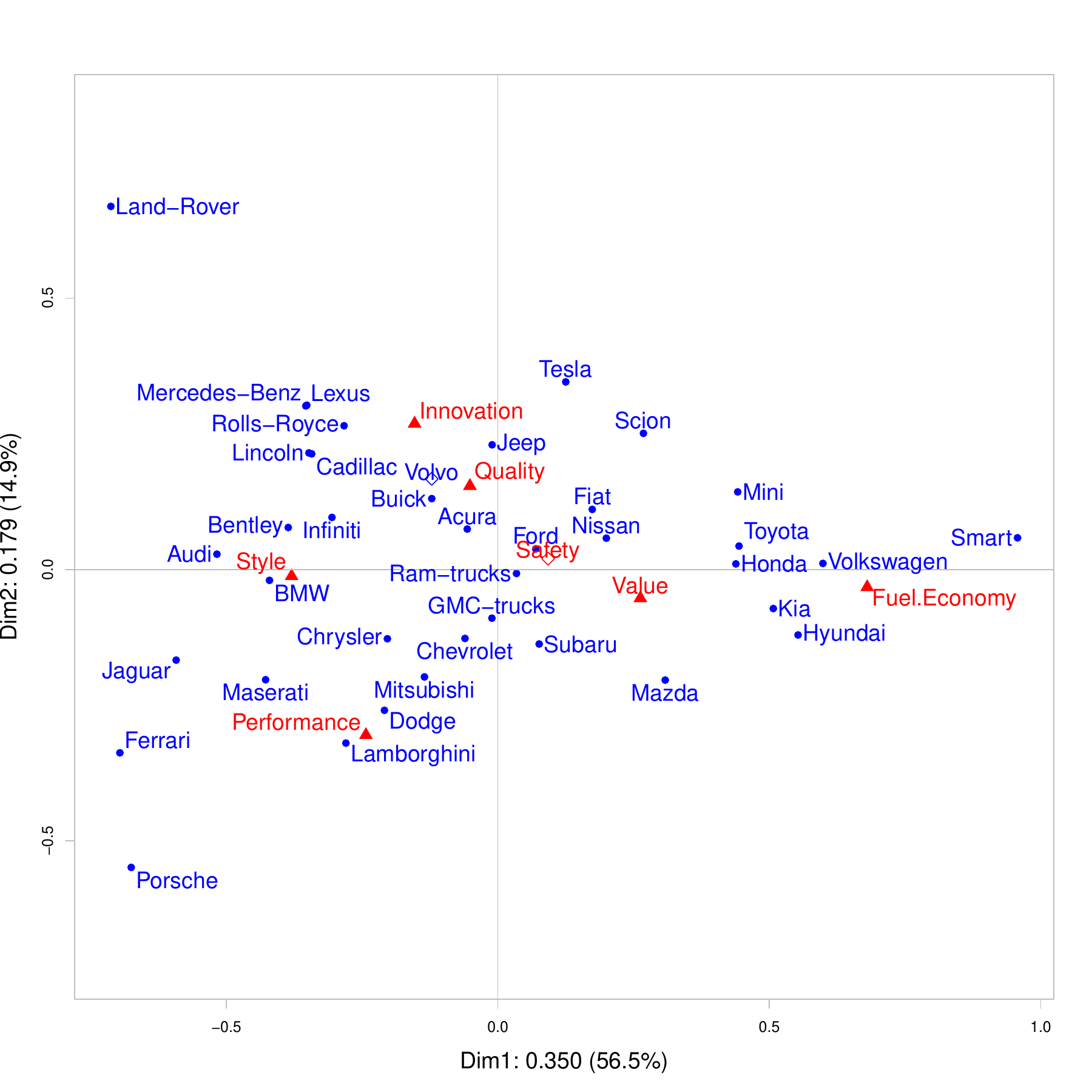}
         \caption{\scriptsize CA supplementary points
Volvo and Safety}\label{F: supbrandscasym}
    \end{subfigure}
    \end{sidewaysfigure}

Figure~\ref{F: brandscasym} shows the symmetric plot of CA. The first four singular values, with percentage of inertia displayed between brackets, are 0.335 (41.3\%), 0.281 (28.9\%), 0.171 (10.7\%), and 0.157 (9.0\%). Using the elbow criterion, we decide to interpret two dimensions.

The first dimension contrasts cars that score high on {\em Fuel Economy} versus cars that score high on {\em Style} and {\em Performance}. On the second dimension the car brand {\em Volvo} is far from other brands of cars, and the attribute {\em Safety} is close by. Where the marginal proportion of {\em Volvo} is 0.024, its contribution to the second dimension is 65.7\%. For {\em Safety} the marginal proportion is 0.132, but the contribution to the second dimension is 75.2\%. In addition, the contribution of cell ({\em Volvo}, {\em Safety}) to the total inertia is 17.7\%. Hence the cell ({\em Volvo}, {\em Safety}) is a cell-wise outlier, leading to outlying points for {\em Volvo} and {\em Safety} on dimension 2.

\begin{table}\footnotesize
\caption{Car data matrix}
        \centering
{% latex table generated in R 4.2.3 by xtable 1.8-4 package
% Tue Jan  2 21:23:53 2024
\begin{tabular}{rrrrrrrrrr}
%\begin{tabular}{m{7em}  m{3em} m{3em}  m{3em}  m{3em} m{3em}  m{3em}  m{3em} | m{3em}  m{3em}}
  \hline
& Fuel Econo. &Innov. & Perform. &Quality & Safety & Style & Value &Total & Proport. \\
%&&&&&&&&&\\
  \hline
Acura & 24 & 38 & 28 & 20 & 28 & 33 & 25 & 196 & 0.017 \\ 
  Audi & 9 & 54 & 54 & 30 & 19 & 67 & 8 & 241 & 0.021 \\ 
  Bentley & 0 & 16 & 18 & 25 & 9 & 27 & 17 & 112 & 0.010 \\ 
  BMW & 14 & 83 & 94 & 55 & 38 & 93 & 35 & 412 & 0.035 \\ 
  Buick & 25 & 48 & 39 & 58 & 52 & 52 & 43 & 317 & 0.027 \\ 
  Cadillac & 14 & 73 & 50 & 76 & 40 & 83 & 36 & 372 & 0.032 \\ 
  Chevrolet & 114 & 103 & 202 & 174 & 140 & 160 & 145 & 1,038 & 0.089 \\ 
  Chrysler & 38 & 65 & 96 & 54 & 54 & 103 & 72 & 482 & 0.041 \\ 
  Dodge & 60 & 61 & 141 & 61 & 63 & 133 & 69 & 588 & 0.050 \\ 
  Ferrari & 0 & 20 & 45 & 10 & 8 & 46 & 5 & 134 & 0.011 \\ 
  Fiat & 19 & 21 & 17 & 20 & 15 & 7 & 16 & 115 & 0.010 \\ 
  Ford & 167 & 180 & 169 & 179 & 161 & 157 & 188 & 1,201 & 0.103 \\ 
  GMC-trucks & 40 & 40 & 64 & 57 & 80 & 50 & 58 & 389 & 0.033 \\ 
  Honda & 163 & 68 & 73 & 118 & 104 & 50 & 135 & 711 & 0.061 \\ 
  Hyundai & 97 & 25 & 31 & 27 & 35 & 42 & 82 & 339 & 0.029 \\ 
  Infiniti & 5 & 39 & 31 & 15 & 10 & 17 & 16 & 133 & 0.011 \\ 
  Jaguar & 0 & 3 & 18 & 19 & 3 & 47 & 12 & 102 & 0.009 \\ 
  Jeep & 18 & 33 & 14 & 51 & 19 & 41 & 52 & 228 & 0.019 \\ 
  Kia & 68 & 30 & 17 & 13 & 24 & 42 & 109 & 303 & 0.026 \\ 
  Lamborghini & 5 & 19 & 37 & 8 & 6 & 23 & 24 & 122 & 0.010 \\ 
  Land-Rover & 0 & 43 & 0 & 5 & 0 & 47 & 2 & 97 & 0.008 \\ 
  Lexus & 10 & 62 & 29 & 50 & 27 & 64 & 26 & 268 & 0.023 \\ 
  Lincoln & 6 & 37 & 23 & 31 & 24 & 40 & 19 & 180 & 0.015 \\ 
  Maserati & 0 & 6 & 9 & 0 & 0 & 41 & 25 & 81 & 0.007 \\ 
  Mazda & 46 & 23 & 34 & 10 & 12 & 26 & 38 & 189 & 0.016 \\ 
  Mercedes-Benz & 8 & 83 & 44 & 87 & 58 & 82 & 42 & 404 & 0.034 \\ 
  Mini & 23 & 12 & 4 & 4 & 13 & 12 & 4 & 72 & 0.006 \\ 
  Mitsubishi & 20 & 13 & 33 & 23 & 7 & 32 & 13 & 141 & 0.012 \\ 
  Nissan & 80 & 68 & 51 & 53 & 52 & 55 & 70 & 429 & 0.037 \\ 
  Porsche & 0 & 17 & 66 & 14 & 6 & 42 & 5 & 150 & 0.013 \\ 
  Ram-trucks & 9 & 22 & 21 & 10 & 18 & 1 & 16 & 97 & 0.008 \\ 
  Rolls-Royce & 0 & 4 & 4 & 35 & 11 & 25 & 17 & 96 & 0.008 \\ 
  Scion & 20 & 24 & 11 & 6 & 11 & 4 & 4 & 80 & 0.007 \\ 
  Smart & 38 & 9 & 3 & 7 & 0 & 5 & 10 & 72 & 0.006 \\ 
  Subaru & 19 & 14 & 32 & 33 & 75 & 20 & 40 & 233 & 0.020 \\ 
  Tesla & 23 & 35 & 10 & 12 & 9 & 15 & 12 & 116 & 0.010 \\ 
  Toyota & 238 & 116 & 95 & 134 & 113 & 74 & 150 & 920 & 0.079 \\ 
  Volkswagen & 90 & 30 & 25 & 37 & 27 & 22 & 46 & 277 & 0.024 \\ 
  Volvo & 9 & 15 & 16 & 31 & 180 & 14 & 11 & 276 & 0.024 \\ 
  \hline
  Total & 1,519 & 1,652 & 1,748 & 1,652 & 1,551 & 1,894 & 1,697&11,713 \\ 
  Proport. & 0.130 & 0.141 & 0.149 & 0.141 & 0.132 & 0.162 & 0.145&& 1.000 \\ 
   \hline
\end{tabular}
}
       \label{T: Tbrands}
\end{table}

\subsubsection{Reconstitution algorithm}\label{Subsub: recalg}

Here we use reconstitution algorithm of order 2 to handle the cell-wise outlier ({\em Volvo}, {\em Safety}). Using the reconstitution algorithm, the value 180 in ({\em Volvo}, {\em Safety}) becomes 27.0. (Reconstitution of order 0 leads an imputed value of 13.1, but the graphic results are similar.) The contribution of cell ({\em Volvo}, {\em Safety})  to
the total inertia went down from 17.7\% to 0.4\%. The first four singular values become 0.334 (51.0\%),  0.186 (15.8\%),  0.170 (13.2\%), and  0.156 (11.1\%). It is clear that the second dimension now is less important, the proportion of inertia went down from 28.9\%  to 15.8\%. The singular values of dimensions 2, 3 and 4 do not differ much, and using the elbow criterion, we decide only to study the first dimension. Also, since in a contingency table the singular value can be interpreted as the canonical correlation between the row variable and the column variable, with 0.186 the second singular value is quite small.

Figure~\ref{F: recbrandscasym} is a symmetric CA plot of the reconstituted table. 
On the first dimension the configuration of row and column points is similar to the configuration of the CA of the original table, except for the change of location of {\em Volvo}. {\em Safety} is still in a similar position, and the reason for this difference between {\em Volvo} and {\em Safety} is that bringing down the value of 180 to 27 has a much larger impact on the profile of {\em Volvo}, that originally had a marginal total of 276, than the marginal total of {\em Safety}, that originally was 1,551. % The row points seem to rotate \todo{how did you get this?} around $30^0$. 
Note that by eliminating the impact of a single cell the new figure is much better readable than Figure~\ref{F: brandscasym}.

By eliminating the influence of a single cell the reconstitution method allows us to arrive at the simple conclusion that (i) there is a single outlying cell for {\em Volvo} and {\em Safety}, as {\em Safety} is chosen as the outstanding characteristic of {\em Volvo} (180 out of 276 scores for {\em Volvo} come from {\em Safety}), and (ii) there is largely a one-dimensional structure for the cars and features going from {\em Land-Rover}, {\em Ferrari} and {\em Porsche} on the left, scoring higher than average on {\em Style} and {\em Performance}, to {\em Smart}, {\em Volkswagen}, {\em Hyundai} and {\em Kia}, on the right, scoring higher on {\em Fuel Economy}, with the other car types and features ordered in between.

\subsubsection{MacroPCA}

We obtain the results of MacroPCA by applying the {\em MacroPCA} function in the R package {\em cellWise} \citep{Raymaekers2023cellWise}. We use the same parameter setting as in  \citet{raymaekers2023challenges} and \citet{Raymaekers2023cellWise},  except for $\alpha = 0.97$ and $k = 2$. By setting $\alpha = 0.97$ we make the number of non-outlying rows as large as possible. We choose $k = 2$ because this simplifies the comparison with the reconstitution of order $h=2$ in CA.

\begin{figure}[H]
\centering
\includegraphics[width=0.27\textwidth, angle =270]
{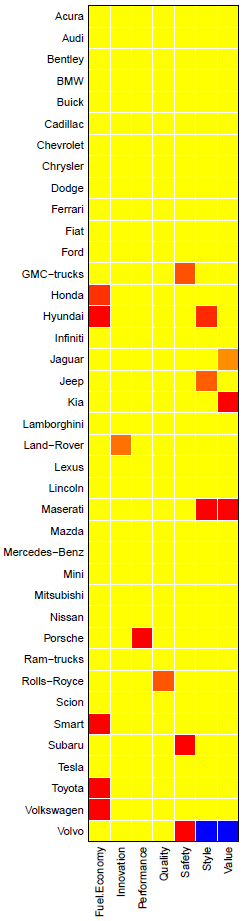}
\caption{Car dataset}\label{F: brandsDDCcelloutlier}
\end{figure}

The results from the first step in MacroPCA, DCC,  provides a cellmap. See Figure~\ref{F: brandsDDCcelloutlier}. The red or blue cells indicate cellwise outliers. Specifically, red cells indicate that the observed values are much larger than the predicted values, and for blue cells the opposite holds. Thus DDC finds 19 cellwise outliers, including the  cellwise outlier  ({\em Volvo}, {\em Safety}) found using the visual inspection employed in Section~\ref{Subsub: recalg}. DDC shows there is no row-wise outlier. However, in the second part of MacroPCA, there are 2 row-wise outliers, which are {\em Land-Rover} and {\em Volvo}.

Figure~\ref{F: macbrandscasym} is the corresponding symmetric CA-type plot.
On the first dimension, the configuration of row and column points is similar to the original Figure~\ref{F: brandscasym}.

\subsubsection{Supplementary points method}

Here we treat {\em Volvo} and {\em Safety} as supplementary points. Thus the table analysed has size $38 \times 6$, and now, row {\em Volvo} and column {\em Safety} have no effect on the solution of CA but are projected into it afterwards. The first four singular values are 0.350 (56.5\%),  0.179 (14.9\%), 0.166 (12.7\%), and 0.150 (10.3\%). 
As in the reconstitution approach, the second dimension is now less important, the proportion of inertia went down from 28.9\% to 14.9\%, and  using the elbow criterion, only the first dimension is to be studied.

Figure~\ref{F: supbrandscasym} shows a symmetric CA plot, where {\em Volvo} and {\em Safety} are added as supplementary points.  
On the first dimension the configuration of row and column points is similar to the original Figure~\ref{F: brandscasym}, except for {\em Volvo}. Again, {\em Safety} is still in a similar position. For this dataset the interpretation using the supplementary points method is very similar to the interpretation using the reconstitution approach.

\subsection{The ocean plastic data}

\noindent The ocean plastic dataset is created by \citet{vonk2024comparative} to analyze how scientific studies on ocean plastic are communicated in press releases. The study analyzed press releases published on EurekAlert! between January 2017 and December 2021. In the analysis, variables defining the four frame elements of \citet{entman1993framing}, namely causal interpretation, problem definition, moral evaluation, and treatment recommendation were noted, resulting in 21 frame variables. Table~\ref{T: fravar} summarizes these framing variables, while a more detailed description can be found in Appendix 1 of \citet{vonk2024comparative}.

The causal interpretation (a) was coded, when the text referred to climate change ({\em CCC}) as a cause of problems.  It was coded whether an entity was held responsible for causing climate change, ocean plastics or related problems ({\em Resp.C.P, Resp.C.I, Resp.C.C, Resp.C.S}, and {\em Resp.C.O}). The problem definition (b) describes different problems ({\em PH, PE, PB, PnB, PT, PC}) or opportunities ({\em OC}) stated in the text. The moral evaluation (d) was coded when an entity was held responsible for solving problems ({\em Resp.T.P, Resp.T.I, Resp.T.C, Resp.T.S}, and {\em Resp.T.O}); when opportunities would be named if problems  were mitigated ({\em OT}); or when the text stated that mitigation of problems was urgently needed ({\em Ur}). The treatment recommendation (c) described a solution that reduced or remedied problems or their cause ({\em Tr}).

\begin{table}\scriptsize
\caption{Frame variables}
\label{T: fravar}
\begin{subtable}[h]{0.5\textwidth}
\centering  
\begin{tabular}{lllllllll}
  \hline
CCC& Cause: Ocean climate change\\
Resp.C.P &Actor responsible for cause: Politics\\
Resp.C.I &Actor responsible for cause: Industry\\
Resp.C.C &Actor responsible for cause: Regions/Countries\\
Resp.C.S &Actor responsible for cause: Society\\
Resp.C.O &Actor responsible for cause: Other\\ 
\hline
\end{tabular}
\caption{Causal interpretation} 
\label{T: Causal interpretation}
    \end{subtable}
\begin{subtable}[h]{0.5\textwidth}
\centering  
\begin{tabular}{lllllllll}
  \hline
PH & Health\\
PE & Economic\\
PB & Biological\\
PnB & Non-Biological\\
PT & Treatment\\
PC & Conflict \\
OC & Opportunity\\
\hline
\end{tabular}
\caption{Problem definition} 
\label{T: Problem definition}
    \end{subtable}
    \begin{subtable}[h]{0.5\textwidth}
\centering  
\begin{tabular}{lllllllll}
  \hline
Tr & Treatment recommendation\\
\hline
\end{tabular}
\caption{Treatment recommendation} 
\label{T: Treatment recommendation}
    \end{subtable}
\begin{subtable}[h]{0.5\textwidth}
\centering  
\begin{tabular}{lllllllll}
  \hline
OT & Opportunity due to treatment\\
Resp.T.P & Actor responsible for treatment: Politics\\
Resp.T.I & Actor responsible for treatment: Industry\\
Resp.T.C & Actor responsible for treatment: Regions/Countries\\
Resp.T.S & Actor responsible for treatment: Society\\
Resp.T.O & Actor responsible for treatment: Other\\
Ur & Urgency to take action\\
\hline
\end{tabular}
\caption{Moral evaluation} 
\label{T: Moral evaluation}
    \end{subtable}
\end{table}

The ocean plastic dataset has 81 press releases in the rows and 21 framing variables in the columns with 0 or 1 in each cell where 1 means the framing variable is present in the text and 0 otherwise. See Table~\ref{T: Topclean} in supplementary materials A. The table has $81 \times 21 = 1,701$ cells of which 1,389 have a value 0. Note that Documents {\em 10}, {\em 34}, {\em 50}, and {\em 81} are identical, and so are Documents {\em 13}, {\em 19}, {\em 26}, {\em 27}, {\em 46}, {\em 56}, {\em 65}, {\em 69}, and {\em 84}, Documents {\em 15}, {\em 71}, and {\em 75}, Documents {\em 17} and {\em 59}, Documents {\em 28} and {\em 31}, Documents {\em 30} and {\em 86}, Documents {\em 41}, {\em 44}, {\em 63}, and {\em 67}, Documents {\em 48} and {\em 77}, and Documents {\em 64}, {\em 72}, and {\em 85}. As the profiles are identical in each group, the points have an identical position in the graphic configurations and we only provide the label {\em 10, 13, 15, 17, 28, 30, 41, 48} and {\em 64}.

\begin{sidewaysfigure}
\caption{Ocean plastic dataset CA plot}\label{F: caopcasym}
\centering
    \begin{subfigure}[b]{0.35\linewidth}
      \includegraphics[width=1\textwidth]{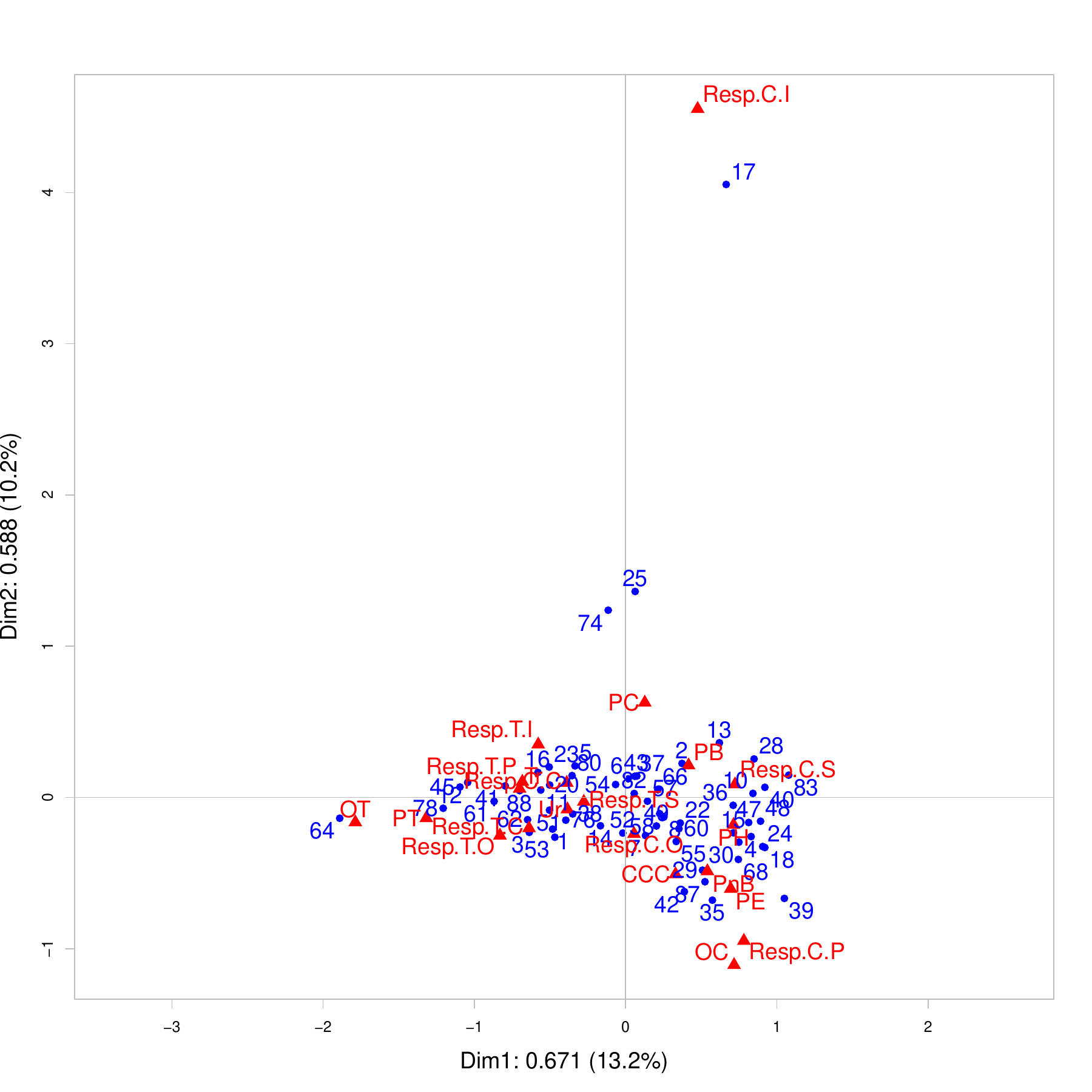}
       \caption{CA original table}\label{F: opcasym}
     \end{subfigure}
    \begin{subfigure}[b]{0.35\linewidth}
      \includegraphics[width=1\textwidth]{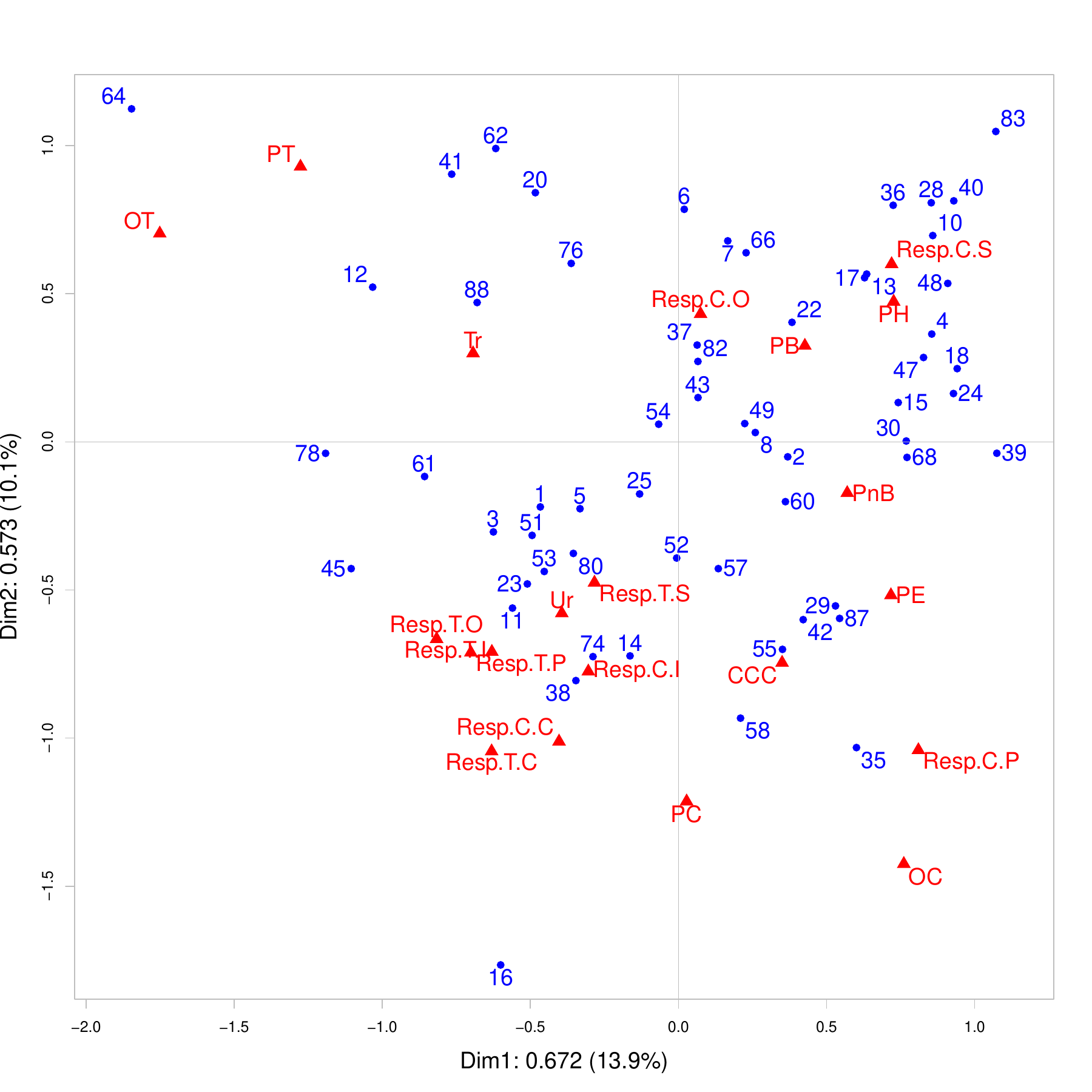}
    \caption{CA reconstitution of order 0 ({\em 17, Resp.C.I}) and ({\em 59, Resp.C.I)}}\label{F: recopcasym}
     \end{subfigure}
    \begin{subfigure}[b]{0.35\linewidth}
      \includegraphics[width=1\textwidth]{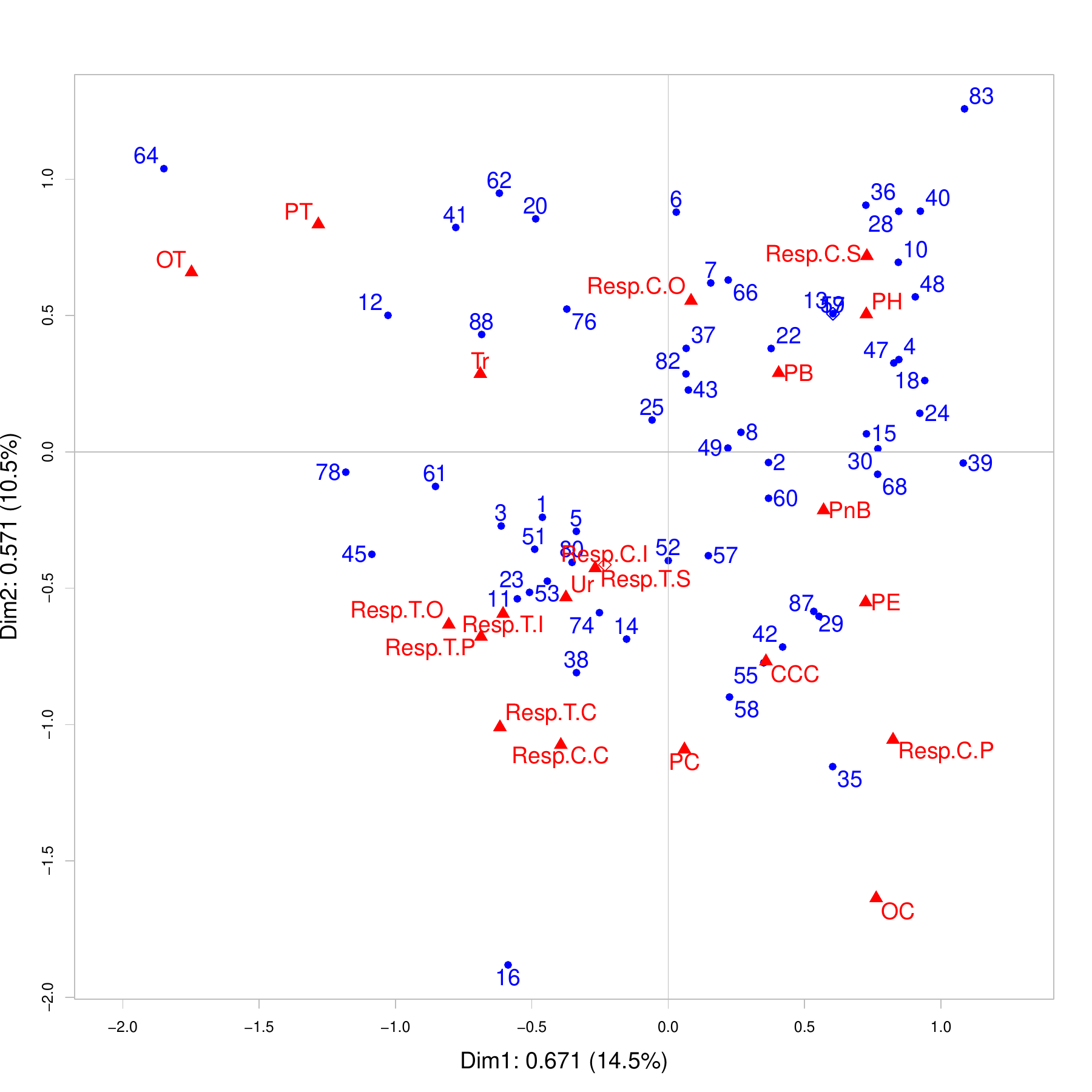}
      \caption{CA supplementary points {\em 17, 59,} and {\em Resp.C.I}}\label{F: supopcasym}
     \end{subfigure}
\end{sidewaysfigure}

Figure~\ref{F: opcasym} is a symmetric plot of the dataset. The first four singular values, with percentages of inertia displayed between brackets, are 0.671 (13.2\%), 0.588 (10.2\%), 0.570 (9.6\%), and 0.544 (8.7\%). The closeness of the singular values shows that the dataset cannot be summarized in a small number of dimensions.

The first dimension contrasts  Opportunity due to treatment ({\em OT}), Treatment related problems ({\em PT}) and Treatment recommendation ({\em Tr}),  Responsibility for treatment framings  {\em T.O, T.P, T.C} and  {\em T.I} on the left versus responsibility for causes framings {\em C.P, C.S} and {\em C.I}, and Problem definitions such as Opportunity ({\em OC}), Health ({\em PH}), Economic ({\em PE}), Non-Biological ({\em PnB}), and Biological ({\em PB}) on the right.
On the second dimension, {\em Resp.C.I}, i.e. industry is responsible for cause, is far from the origin. The marginal proportion of {\em Resp.C.I} is 0.013, and its contribution to the second dimension is 76.9\%. {\em Resp.C.I} masks the visualisation of the structure in the  dataset and reduces the readability of this map. Documents {\em 17, 59}, which have identical scores, are far from the origin and are closest to  {\em Resp.C.I}. The marginal proportion of documents {\em 17/59} jointly is 0.013, yet its contribution to the second dimension is 61.0\%. Also, the contribution of the two cells ({\em 17/59}, {\em Resp.C.I})  to the total inertia is 7.0\%, which is large (note that there are $81 \times 21$ cells). Hence the cells ({\em 17/59}, {\em Resp.C.I}) are cell-wise outliers, leading to outlying points
for {\em 17, 59} and {\em Resp.C.I} on dimension 2.

\subsubsection{Reconstitution algorithm}

Again, we used the reconstitution algorithm of order 2 to handle the cell-wise outliers. However, this created a negative imputed value $-0.0006$ for outlying cells ({\em 17/59, Resp.C.I}). Negative values are not easy to interpret in an incidence matrix. Therefore, we applied reconstitution of order 0. This yields value 0.0065 for the cells ({\em 17/59, Resp.C.I}). Now the documents {\em 17/59}, having a 1 in the framing variable {\em PB}, 0.0065 in Resp.C.I and otherwise 0, are similar to documents {\em 13, 19, 26, 27, 46, 56, 65, 69, 84} which have 1 in {\em PB} and 0 otherwise. The first four singular values are 0.672 (13.9\%), 0.573 (10.1\%), 0.548 (9.3\%), and 0.519 (8.3\%).

Figure~\ref{F: recopcasym} is a symmetric CA plot of the reconstituted table. On the first dimension the configuration of column points is similar to the configuration in Figure~\ref{F: opcasym}, except for {\em Resp.C.I}. {\em Resp.C.I} is not close to Documents {\em 17/59}, and {\em Resp.C.I, 17/59} are not far from the origin. 
Now, the contributions to the second dimension of {\em Resp.C.I} is only 1.2\% and of {\em 17/59} jointly 0.6\%.

By reducing the influence of cells ({\em 17/59, Resp.C.I}), the new figure is much better readable than Figure~\ref{F: opcasym}. A full interpretation of the table makes use of the  outliers found in standard CA, and the CA solution found with the reconstitution method. The standard CA reveals a strong positive relation between {\em 17/59} and {\em Resp.C.I}. We interpret the CA solution found with the reconstitution method by interpreting the four quadrants of Figure~\ref{F: recopcasym}.
\begin{itemize}
    \item Press releases in the first quadrant focus on problems related to biology ({\em PB}) and human health ({\em PH}), and place the responsibility for causes at society ({\em Resp.C.S}); 
    \item The second quadrant represents problems related to treatment ({\em PT}) and solutions to these problems in the form of treatment ({\em Tr}), and opportunity if treatment is carried out ({\em OT});
    \item Press releases in the third quadrant focus on the urgency to treat ocean plastic ({\em Ur}) and hold entity responsible for carrying out that treatment ({\em Resp.T.C, Resp.T.P, Resp.T.I, Resp.T.O, Resp.T.S}). In some cases they also state the responsibility for cause at industry ({\em Resp.C.I}) and specific regions/countries ({\em Resp.C.C}); 
    \item Press releases in the fourth quadrant focus on the interconnections between ocean plastic and climate change ({\em CCC}) and they state non-biological ({\em PnB}) and economic consequences ({\em PE}). The fourth quadrant also represents the responsibility for cause at politics ({\em Resp.C.P}) and opportunity due to problems ({\em OC}). We note that the marginal frequencies of {\em Resp.C.P} and {\em OC} are low, namely 1 and 2 respectively. 
\end{itemize}

\subsubsection{Supplementary points method}

Here we treat {\em 17/59} and {\em Resp.C.I} as supplementary points. Thus the size of the table analysed is 79 $\times$ 20. Now, due to deleting {\em Resp.C.I}, documents {\em 25} and {\em 54} are also identical. Rows {17/59} and column {\em Resp.C.I} have no effect on the solution of CA but are projected into it afterwards. The first four singular values are 0.671 (14.5\%), 0.571 (10.5\%), 0.544 (9.6\%), and 0.511 (8.4\%).

Figure~\ref{F: supopcasym} shows the symmetric CA plot for the supplementary points method. Figure~\ref{F: supopcasym} is similar to Figure~\ref{F: recopcasym}.

\section{Discussion and conclusion}\label{S: con}

In this paper, we propose to use the reconstitution algorithm of order $h$ to deal with outlying cells in CA. The reconstitution algorithm of order $h$ can reduce the effects of single outlying cells on the CA solution. We compare it with MacroPCA and the supplementary points method. 

In comparison to the reconstitution approach, MacroPCA imputes outlying cells in the matrix of standardized residuals instead of in the original matrix. Apart from imputing cell-wise outliers, it can also eliminate complete rows. Yet, MacroPCA is not as transparent and straightforward as the reconstitution approach. One of the reasons is that it is originally proposed for the analysis continuous data and makes distributional assumptions, which do not hold for the reconstitution approach. Due to these distributional assumptions, that do not always fit with how the data originate, in our view it appears to flag too many cells as outlying cells. 

The supplementary points method deletes  complete rows or columns. In contrast, the reconstitution algorithm only reduces the influence of outlying cells. Thus, the reconstitution algorithm uses more information in the data and is, from this perspective, preferable.

We analysed two real data sets to illustrate the use of the reconstitution algorithm and compared the algorithm with the supplementary points method and MacroPCA. For the contingency table car dataset, the three methods yielded similar results. For the ocean plastic dataset, the reconstitution algorithm and the supplementary points method had similar results, but MacroPCA failed.

We are not able to show empirically that the reconstitution method is preferable over the supplementary points method and MacroPCA. However, on theoretical grounds the reconstitution method is preferable: it eliminates only single cells to handle outlier problems, thus it is not necessarily deleting more information than is necessary.

\section{Software}\label{S: sof}

The reconstitution algorithm of order $h$ is implemented by a function {\em reconca}  both for $h = 0$ and $h > 0$. The function is written by adjusting the function {\em imputeCA} in the R Package {\em missMDA}. \citet{josse2016missmda} proposed the R package {\em missMDA} for handling missing values in multivariate data analysis, where the function {\em imputeCA} is meant for missing values in CA. Another R Package, which can perform a reconstitution algorithm of order zero, is {\em anacor}, proposed for simple and canonical CA by \citet{de2009simple}, to deal with missing data in CA. 

The MacroPCA method is performed by the {\em MacroPCA} function in R package {\em cellWise}. The MacroPCA method is proposed for PCA \citep{hubert2019macropca} and adjusted for CA \citep{raymaekers2023challenges}.  To fit CA, the original matrix is replaced with the matrix of standardized residuals.

The code to reproduce the results of this paper including the function {\em reconca} is on the GitHub website \url{https://github.com/qianqianqi28/ca-outlier}.

\section*{Acknowledgement}

Author Qianqian Qi is supported by the China Scholarship Council (CSC202007720017).

\bibliography{references.bib}

\appendixqq

\appendixqqsection{Ocean plastic dataset of  of size $81 \times 21$}\label{SS: opclean}

\begin{center}
{\setlength\tabcolsep{1.5pt}\tiny
{\begin{longtable}{rrrrrrrrrrrrrrrrrrrrrr|rr}
\caption{Ocean plastic data matrix}
\label{T: Topclean}
\\
\hline & PH & PE & PB & PnB & PT & PC & OC & OT & CCC & Resp.C.P & Resp.C.I & Resp.C.C & Resp.C.S & Resp.C.O & Resp.T.P & Resp.T.I & Resp.T.C & Resp.T.S & Resp.T.O & Ur & Tr &Total&Proport.\\ 
\hline 
\endfirsthead
\multicolumn{24}{r}%
\centering
{{ \tablename\ \thetable{} continued from previous page}} \\
\hline & PH & PE & PB & PnB & PT & PC & OC & OT & CCC & Resp.C.P & Resp.C.I & Resp.C.C & Resp.C.S & Resp.C.O & Resp.T.P & Resp.T.I & Resp.T.C & Resp.T.S & Resp.T.O & Ur & Tr &Total &Proportion  \\ 
\hline 
\endhead

\hline 
\endfoot

1 & 0 & 0 & 1 & 1 & 0 & 0 & 0 & 1 & 1 & 0 & 0 & 0 & 0 & 0 & 1 & 0 & 0 & 0 & 0 & 1 & 1 & 7 & 0.022 \\ 
  2 & 0 & 0 & 1 & 0 & 0 & 0 & 0 & 0 & 0 & 0 & 0 & 1 & 1 & 0 & 0 & 0 & 0 & 0 & 0 & 0 & 0 & 3 & 0.010 \\ 
  3 & 1 & 0 & 0 & 1 & 1 & 0 & 0 & 1 & 1 & 0 & 0 & 0 & 0 & 1 & 1 & 1 & 1 & 1 & 1 & 1 & 1 & 13 & 0.042 \\ 
  4 & 1 & 0 & 1 & 1 & 0 & 0 & 0 & 0 & 0 & 0 & 0 & 0 & 0 & 0 & 0 & 0 & 0 & 0 & 0 & 0 & 0 & 3 & 0.010 \\ 
  5 & 0 & 0 & 1 & 0 & 0 & 0 & 0 & 0 & 0 & 0 & 0 & 1 & 0 & 0 & 0 & 0 & 0 & 0 & 0 & 0 & 1 & 3 & 0.010 \\ 
  6 & 0 & 0 & 0 & 0 & 0 & 0 & 0 & 0 & 0 & 0 & 0 & 0 & 1 & 0 & 0 & 0 & 0 & 0 & 0 & 0 & 1 & 2 & 0.006 \\ 
  7 & 1 & 0 & 1 & 1 & 1 & 0 & 0 & 0 & 0 & 0 & 0 & 0 & 0 & 0 & 0 & 0 & 0 & 0 & 0 & 0 & 0 & 4 & 0.013 \\ 
  8 & 1 & 1 & 1 & 0 & 0 & 0 & 0 & 0 & 0 & 0 & 0 & 0 & 1 & 0 & 0 & 0 & 0 & 1 & 0 & 1 & 1 & 7 & 0.022 \\ 
  10 & 1 & 0 & 1 & 0 & 0 & 0 & 0 & 0 & 0 & 0 & 0 & 0 & 0 & 0 & 0 & 0 & 0 & 0 & 0 & 0 & 0 & 2 & 0.006 \\ 
  11 & 0 & 0 & 1 & 0 & 0 & 0 & 0 & 0 & 0 & 0 & 0 & 0 & 0 & 0 & 1 & 0 & 1 & 1 & 0 & 0 & 1 & 5 & 0.016 \\ 
  12 & 0 & 0 & 0 & 0 & 0 & 0 & 0 & 0 & 0 & 0 & 0 & 0 & 0 & 0 & 0 & 0 & 0 & 0 & 0 & 0 & 1 & 1 & 0.003 \\ 
  13 & 0 & 0 & 1 & 0 & 0 & 0 & 0 & 0 & 0 & 0 & 0 & 0 & 0 & 0 & 0 & 0 & 0 & 0 & 0 & 0 & 0 & 1 & 0.003 \\ 
  14 & 1 & 0 & 1 & 1 & 0 & 0 & 0 & 0 & 1 & 0 & 0 & 0 & 0 & 0 & 1 & 1 & 1 & 1 & 1 & 0 & 0 & 9 & 0.029 \\ 
  15 & 0 & 0 & 1 & 1 & 0 & 0 & 0 & 0 & 0 & 0 & 0 & 0 & 0 & 0 & 0 & 0 & 0 & 0 & 0 & 0 & 0 & 2 & 0.006 \\ 
  16 & 0 & 0 & 0 & 0 & 0 & 0 & 0 & 0 & 0 & 0 & 0 & 1 & 0 & 0 & 0 & 0 & 0 & 0 & 0 & 0 & 0 & 1 & 0.003 \\ 
  17 & 0 & 0 & 1 & 0 & 0 & 0 & 0 & 0 & 0 & 0 & 1 & 0 & 0 & 0 & 0 & 0 & 0 & 0 & 0 & 0 & 0 & 2 & 0.006 \\ 
  18 & 1 & 1 & 1 & 1 & 0 & 0 & 0 & 0 & 0 & 0 & 0 & 0 & 1 & 0 & 0 & 0 & 0 & 0 & 0 & 0 & 0 & 5 & 0.016 \\ 
  19 & 0 & 0 & 1 & 0 & 0 & 0 & 0 & 0 & 0 & 0 & 0 & 0 & 0 & 0 & 0 & 0 & 0 & 0 & 0 & 0 & 0 & 1 & 0.003 \\ 
  20 & 0 & 0 & 1 & 0 & 0 & 0 & 0 & 1 & 0 & 0 & 0 & 0 & 1 & 0 & 0 & 0 & 0 & 0 & 0 & 0 & 1 & 4 & 0.013 \\ 
  22 & 1 & 0 & 1 & 1 & 0 & 0 & 0 & 0 & 0 & 0 & 0 & 0 & 0 & 0 & 0 & 0 & 0 & 0 & 0 & 0 & 1 & 4 & 0.013 \\ 
  23 & 0 & 0 & 1 & 0 & 0 & 0 & 0 & 0 & 0 & 0 & 0 & 1 & 0 & 0 & 1 & 0 & 0 & 0 & 0 & 0 & 1 & 4 & 0.013 \\ 
  24 & 1 & 1 & 1 & 0 & 0 & 0 & 0 & 0 & 0 & 0 & 0 & 0 & 0 & 0 & 0 & 0 & 0 & 0 & 0 & 0 & 0 & 3 & 0.010 \\ 
  25 & 0 & 0 & 1 & 0 & 0 & 0 & 0 & 0 & 0 & 0 & 1 & 0 & 1 & 0 & 0 & 0 & 0 & 1 & 0 & 1 & 1 & 6 & 0.019 \\ 
  26 & 0 & 0 & 1 & 0 & 0 & 0 & 0 & 0 & 0 & 0 & 0 & 0 & 0 & 0 & 0 & 0 & 0 & 0 & 0 & 0 & 0 & 1 & 0.003 \\ 
  27 & 0 & 0 & 1 & 0 & 0 & 0 & 0 & 0 & 0 & 0 & 0 & 0 & 0 & 0 & 0 & 0 & 0 & 0 & 0 & 0 & 0 & 1 & 0.003 \\ 
  28 & 0 & 0 & 1 & 0 & 0 & 0 & 0 & 0 & 0 & 0 & 0 & 0 & 1 & 0 & 0 & 0 & 0 & 0 & 0 & 0 & 0 & 2 & 0.006 \\ 
  29 & 0 & 1 & 1 & 1 & 0 & 0 & 0 & 0 & 1 & 0 & 0 & 0 & 0 & 0 & 0 & 0 & 0 & 1 & 0 & 0 & 0 & 5 & 0.016 \\ 
  30 & 0 & 0 & 1 & 1 & 0 & 0 & 0 & 0 & 1 & 0 & 0 & 0 & 1 & 0 & 0 & 0 & 0 & 0 & 0 & 0 & 0 & 4 & 0.013 \\ 
  31 & 0 & 0 & 1 & 0 & 0 & 0 & 0 & 0 & 0 & 0 & 0 & 0 & 1 & 0 & 0 & 0 & 0 & 0 & 0 & 0 & 0 & 2 & 0.006 \\ 
  34 & 1 & 0 & 1 & 0 & 0 & 0 & 0 & 0 & 0 & 0 & 0 & 0 & 0 & 0 & 0 & 0 & 0 & 0 & 0 & 0 & 0 & 2 & 0.006 \\ 
  35 & 0 & 1 & 1 & 1 & 0 & 0 & 1 & 0 & 1 & 0 & 0 & 1 & 0 & 0 & 0 & 0 & 0 & 0 & 0 & 0 & 0 & 6 & 0.019 \\ 
  36 & 1 & 0 & 1 & 0 & 0 & 0 & 0 & 0 & 0 & 0 & 0 & 0 & 1 & 1 & 0 & 0 & 0 & 0 & 0 & 0 & 0 & 4 & 0.013 \\ 
  37 & 0 & 0 & 1 & 0 & 0 & 0 & 0 & 0 & 0 & 0 & 0 & 0 & 1 & 0 & 0 & 0 & 0 & 1 & 0 & 0 & 1 & 4 & 0.013 \\ 
  38 & 0 & 0 & 1 & 0 & 0 & 0 & 0 & 0 & 1 & 0 & 0 & 1 & 0 & 0 & 0 & 0 & 1 & 1 & 0 & 1 & 1 & 7 & 0.022 \\ 
  39 & 1 & 1 & 0 & 0 & 0 & 0 & 0 & 0 & 0 & 0 & 0 & 0 & 0 & 0 & 0 & 0 & 0 & 0 & 0 & 0 & 0 & 2 & 0.006 \\ 
  40 & 1 & 0 & 1 & 0 & 0 & 0 & 0 & 0 & 0 & 0 & 0 & 0 & 1 & 0 & 0 & 0 & 0 & 0 & 0 & 0 & 0 & 3 & 0.010 \\ 
  41 & 0 & 0 & 1 & 0 & 1 & 0 & 0 & 0 & 0 & 0 & 0 & 0 & 0 & 0 & 0 & 0 & 0 & 0 & 0 & 0 & 1 & 3 & 0.010 \\ 
  42 & 0 & 0 & 1 & 1 & 0 & 0 & 1 & 0 & 1 & 0 & 0 & 0 & 0 & 0 & 0 & 0 & 0 & 0 & 0 & 0 & 1 & 5 & 0.016 \\ 
  43 & 1 & 0 & 1 & 0 & 0 & 0 & 0 & 0 & 0 & 0 & 0 & 0 & 1 & 0 & 0 & 1 & 0 & 1 & 0 & 0 & 1 & 6 & 0.019 \\ 
  44 & 0 & 0 & 1 & 0 & 1 & 0 & 0 & 0 & 0 & 0 & 0 & 0 & 0 & 0 & 0 & 0 & 0 & 0 & 0 & 0 & 1 & 3 & 0.010 \\ 
  45 & 0 & 0 & 0 & 0 & 0 & 0 & 0 & 1 & 0 & 0 & 0 & 0 & 0 & 0 & 1 & 1 & 0 & 1 & 0 & 1 & 1 & 6 & 0.019 \\ 
  46 & 0 & 0 & 1 & 0 & 0 & 0 & 0 & 0 & 0 & 0 & 0 & 0 & 0 & 0 & 0 & 0 & 0 & 0 & 0 & 0 & 0 & 1 & 0.003 \\ 
  47 & 1 & 0 & 1 & 0 & 0 & 0 & 0 & 0 & 1 & 0 & 0 & 0 & 1 & 0 & 0 & 0 & 0 & 0 & 0 & 0 & 0 & 4 & 0.013 \\ 
  48 & 1 & 0 & 1 & 1 & 0 & 0 & 0 & 0 & 0 & 0 & 0 & 0 & 1 & 0 & 0 & 0 & 0 & 0 & 0 & 0 & 0 & 4 & 0.013 \\ 
  49 & 0 & 1 & 1 & 0 & 0 & 0 & 0 & 0 & 0 & 0 & 0 & 0 & 0 & 0 & 0 & 0 & 0 & 0 & 0 & 0 & 1 & 3 & 0.010 \\ 
  50 & 1 & 0 & 1 & 0 & 0 & 0 & 0 & 0 & 0 & 0 & 0 & 0 & 0 & 0 & 0 & 0 & 0 & 0 & 0 & 0 & 0 & 2 & 0.006 \\ 
  51 & 0 & 1 & 1 & 0 & 1 & 0 & 0 & 0 & 0 & 0 & 0 & 1 & 0 & 0 & 1 & 0 & 0 & 0 & 0 & 1 & 1 & 7 & 0.022 \\ 
  52 & 0 & 0 & 1 & 1 & 0 & 0 & 0 & 0 & 1 & 0 & 0 & 0 & 0 & 0 & 0 & 0 & 0 & 1 & 0 & 1 & 1 & 6 & 0.019 \\ 
  53 & 0 & 0 & 0 & 1 & 1 & 0 & 0 & 0 & 1 & 0 & 0 & 1 & 0 & 0 & 0 & 0 & 0 & 1 & 0 & 1 & 1 & 7 & 0.022 \\ 
  54 & 0 & 0 & 1 & 0 & 0 & 0 & 0 & 0 & 0 & 0 & 0 & 0 & 1 & 0 & 0 & 0 & 0 & 1 & 0 & 1 & 1 & 5 & 0.016 \\ 
  55 & 0 & 0 & 1 & 1 & 0 & 0 & 0 & 0 & 1 & 0 & 0 & 1 & 0 & 0 & 0 & 0 & 0 & 0 & 0 & 0 & 0 & 4 & 0.013 \\ 
  56 & 0 & 0 & 1 & 0 & 0 & 0 & 0 & 0 & 0 & 0 & 0 & 0 & 0 & 0 & 0 & 0 & 0 & 0 & 0 & 0 & 0 & 1 & 0.003 \\ 
  57 & 0 & 0 & 1 & 1 & 0 & 1 & 0 & 0 & 1 & 0 & 0 & 0 & 1 & 0 & 0 & 0 & 0 & 1 & 0 & 1 & 1 & 8 & 0.026 \\ 
  58 & 0 & 1 & 1 & 0 & 0 & 1 & 0 & 0 & 1 & 0 & 0 & 0 & 0 & 0 & 0 & 0 & 0 & 1 & 0 & 1 & 0 & 6 & 0.019 \\ 
  59 & 0 & 0 & 1 & 0 & 0 & 0 & 0 & 0 & 0 & 0 & 1 & 0 & 0 & 0 & 0 & 0 & 0 & 0 & 0 & 0 & 0 & 2 & 0.006 \\ 
  60 & 1 & 0 & 1 & 1 & 0 & 0 & 0 & 0 & 1 & 0 & 0 & 0 & 1 & 0 & 1 & 0 & 0 & 0 & 0 & 1 & 0 & 7 & 0.022 \\ 
  61 & 0 & 0 & 1 & 0 & 1 & 0 & 0 & 0 & 0 & 0 & 0 & 0 & 0 & 0 & 1 & 0 & 0 & 0 & 1 & 1 & 1 & 6 & 0.019 \\ 
  62 & 1 & 0 & 0 & 0 & 1 & 0 & 0 & 0 & 0 & 0 & 0 & 0 & 0 & 0 & 0 & 0 & 0 & 0 & 0 & 0 & 1 & 3 & 0.010 \\ 
  63 & 0 & 0 & 1 & 0 & 1 & 0 & 0 & 0 & 0 & 0 & 0 & 0 & 0 & 0 & 0 & 0 & 0 & 0 & 0 & 0 & 1 & 3 & 0.010 \\ 
  64 & 0 & 0 & 0 & 0 & 1 & 0 & 0 & 1 & 0 & 0 & 0 & 0 & 0 & 0 & 0 & 0 & 0 & 0 & 0 & 0 & 1 & 3 & 0.010 \\ 
  65 & 0 & 0 & 1 & 0 & 0 & 0 & 0 & 0 & 0 & 0 & 0 & 0 & 0 & 0 & 0 & 0 & 0 & 0 & 0 & 0 & 0 & 1 & 0.003 \\ 
  66 & 1 & 0 & 1 & 0 & 0 & 0 & 0 & 0 & 0 & 0 & 0 & 0 & 0 & 0 & 0 & 0 & 0 & 0 & 0 & 0 & 1 & 3 & 0.010 \\ 
  67 & 0 & 0 & 1 & 0 & 1 & 0 & 0 & 0 & 0 & 0 & 0 & 0 & 0 & 0 & 0 & 0 & 0 & 0 & 0 & 0 & 1 & 3 & 0.010 \\ 
  68 & 1 & 0 & 1 & 1 & 0 & 0 & 0 & 0 & 1 & 0 & 0 & 0 & 0 & 0 & 0 & 0 & 0 & 0 & 0 & 0 & 0 & 4 & 0.013 \\ 
  69 & 0 & 0 & 1 & 0 & 0 & 0 & 0 & 0 & 0 & 0 & 0 & 0 & 0 & 0 & 0 & 0 & 0 & 0 & 0 & 0 & 0 & 1 & 0.003 \\ 
  71 & 0 & 0 & 1 & 1 & 0 & 0 & 0 & 0 & 0 & 0 & 0 & 0 & 0 & 0 & 0 & 0 & 0 & 0 & 0 & 0 & 0 & 2 & 0.006 \\ 
  72 & 0 & 0 & 0 & 0 & 1 & 0 & 0 & 1 & 0 & 0 & 0 & 0 & 0 & 0 & 0 & 0 & 0 & 0 & 0 & 0 & 1 & 3 & 0.010 \\ 
  74 & 1 & 0 & 1 & 0 & 0 & 1 & 0 & 0 & 0 & 0 & 1 & 1 & 0 & 0 & 1 & 1 & 0 & 0 & 0 & 0 & 1 & 8 & 0.026 \\ 
  75 & 0 & 0 & 1 & 1 & 0 & 0 & 0 & 0 & 0 & 0 & 0 & 0 & 0 & 0 & 0 & 0 & 0 & 0 & 0 & 0 & 0 & 2 & 0.006 \\ 
  76 & 0 & 0 & 1 & 1 & 1 & 0 & 0 & 0 & 0 & 0 & 0 & 0 & 0 & 0 & 0 & 0 & 0 & 0 & 0 & 0 & 1 & 4 & 0.013 \\ 
  77 & 1 & 0 & 1 & 1 & 0 & 0 & 0 & 0 & 0 & 0 & 0 & 0 & 1 & 0 & 0 & 0 & 0 & 0 & 0 & 0 & 0 & 4 & 0.013 \\ 
  78 & 0 & 0 & 0 & 0 & 1 & 0 & 0 & 1 & 0 & 0 & 0 & 1 & 0 & 0 & 0 & 0 & 0 & 1 & 0 & 1 & 1 & 6 & 0.019 \\ 
  80 & 0 & 0 & 1 & 0 & 0 & 0 & 0 & 0 & 0 & 0 & 0 & 1 & 0 & 0 & 0 & 0 & 0 & 1 & 0 & 0 & 1 & 4 & 0.013 \\ 
  81 & 1 & 0 & 1 & 0 & 0 & 0 & 0 & 0 & 0 & 0 & 0 & 0 & 0 & 0 & 0 & 0 & 0 & 0 & 0 & 0 & 0 & 2 & 0.006 \\ 
  82 & 1 & 0 & 1 & 0 & 0 & 0 & 0 & 0 & 0 & 0 & 0 & 0 & 0 & 0 & 0 & 0 & 0 & 1 & 0 & 0 & 1 & 4 & 0.013 \\ 
  83 & 0 & 0 & 0 & 0 & 0 & 0 & 0 & 0 & 0 & 0 & 0 & 0 & 1 & 0 & 0 & 0 & 0 & 0 & 0 & 0 & 0 & 1 & 0.003 \\ 
  84 & 0 & 0 & 1 & 0 & 0 & 0 & 0 & 0 & 0 & 0 & 0 & 0 & 0 & 0 & 0 & 0 & 0 & 0 & 0 & 0 & 0 & 1 & 0.003 \\ 
  85 & 0 & 0 & 0 & 0 & 1 & 0 & 0 & 1 & 0 & 0 & 0 & 0 & 0 & 0 & 0 & 0 & 0 & 0 & 0 & 0 & 1 & 3 & 0.010 \\ 
  86 & 0 & 0 & 1 & 1 & 0 & 0 & 0 & 0 & 1 & 0 & 0 & 0 & 1 & 0 & 0 & 0 & 0 & 0 & 0 & 0 & 0 & 4 & 0.013 \\ 
  87 & 1 & 1 & 1 & 1 & 0 & 0 & 0 & 0 & 1 & 1 & 0 & 0 & 0 & 0 & 0 & 0 & 0 & 1 & 0 & 1 & 0 & 8 & 0.026 \\ 
  88 & 0 & 0 & 1 & 0 & 1 & 0 & 0 & 0 & 0 & 0 & 0 & 0 & 0 & 0 & 0 & 0 & 0 & 1 & 0 & 0 & 1 & 4 & 0.013 \\ 
  \hline
   Total & 26 & 10 & 68 & 25 & 16 & 3 & 2 & 8 & 18 & 1 & 4 & 12 & 21 & 2 & 10 & 5 & 4 & 20 & 3 & 16 & 38 & 312 \\ 
  Proport. & 0.083 & 0.032 & 0.218 & 0.080 & 0.051 & 0.010 & 0.006 & 0.026 & 0.058 & 0.003 & 0.013 & 0.038 & 0.067 & 0.006 & 0.032 & 0.016 & 0.013 & 0.064 & 0.010 & 0.051 & 0.122 & & 1.000 \\ 
  \hline
\end{longtable}}
}
\end{center}

\appendixqqsection{MacroPCA for ocean plastic dataset}

\begin{figure}[H]
\centering
\includegraphics[width=0.27\textwidth, angle =270]
{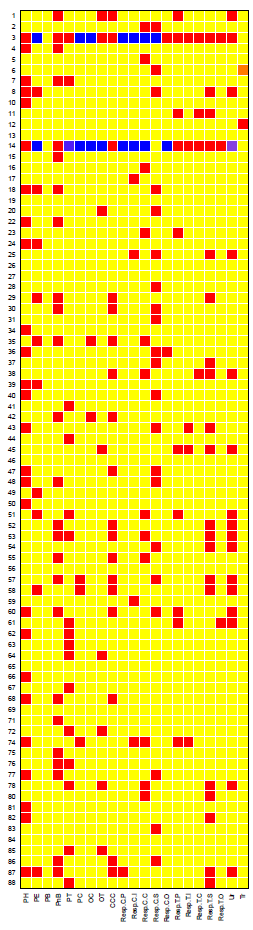}
\caption{Ocean plastic dataset}\label{F: opDDCcelloutlier}
\end{figure}

We use the same parameter as in the car dataset to deal with the ocean plastic dataset. Figure~\ref{F: opDDCcelloutlier} is a cellmap based on the results from the first part DDC of MacroPCA. MacroPCA flags 225 cellwise outliers. The 225 cellwise outliers include ({\em 17}, {\em Resp.C.I}) and ({\em 59}, {\em Resp.C.I}), which we took as cellwise outliers in reconstitution of order $h$. 
  
\begin{figure}[H]
            \centering
            \includegraphics[width=0.6\textwidth]{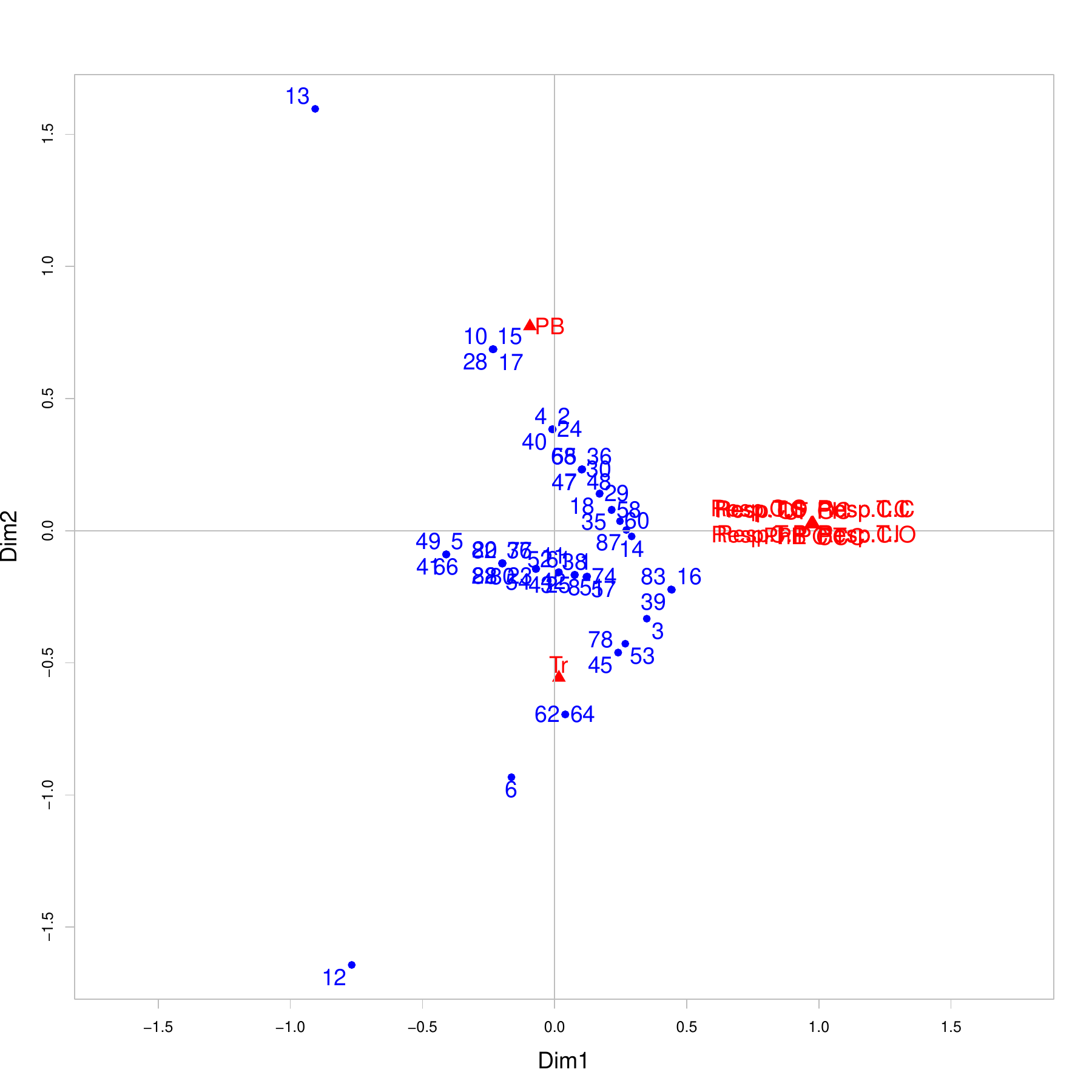}
    \caption{CA MacroPCA of order 2}\label{F: macopcasym}
        \end{figure}

Figure~\ref{F: macopcasym} is the corresponding symmetric CA-type plot based on MacroPCA. MacroPCA does not work well in the ocean plastic dataset. The reason may be that the analyzed matrix severely violates the assumption of multivariate Gaussian distribution.

\end{document}